\documentclass[aps,pra,showpacs,reprint,superscriptaddress]{revtex4-1}

\usepackage{graphicx}
\usepackage{amsmath,amssymb,bm}
\usepackage{siunitx}
\usepackage[utf8]{inputenc}
\usepackage[francais,british]{babel}    
\def\idmx{1\!{\rm l}}
\begin{document}


\title{Fermi golden rule beyond the Zeno regime}


\author{Vincent Debierre}
\affiliation{Aix Marseille Université, CNRS, École Centrale de Marseille, Institut Fresnel UMR 7249, 13013 Marseille, France.}
\author{Isabelle Goessens}
\affiliation{Vrije Universiteit Brussel, Pleinlaan 2, B-1050 Elsene, Belgium.}
\author{Edouard Brainis}
\email[]{Edouard.Brainis@UGent.be}
\affiliation{Physics and Chemistry of Nanostructures, Ghent University, Krijgslaan 281, B-9000 Gent, Belgium}
\affiliation{Centre of Nano- and Biophotonics (NB-Photonics), Ghent University, Sint-Pietersnieuwstraat 41, Ghent B-9000, Belgium}
\author{Thomas Durt}
\affiliation{Aix Marseille Université, CNRS, École Centrale de Marseille, Institut Fresnel UMR 7249, 13013 Marseille, France.}


\date{\today}

\begin{abstract}
We reconsider the problem of the spontaneous emission of light by an excited atomic state. We scrutinize the survival probability of this excited state for very short times, in the so-called Zeno regime, for which we show that the dynamics is dictated by a coherent --in phase -- response of the on-shell and off-shell vacuum modes.
We also develop a perturbative approach in order to interpolate between different temporal regimes: the Zeno, golden rule (linear) and Wigner-Weisskopf (exponential) regimes. We compare results obtained with the $\hat{\mathbf{E}}\cdot\hat{\mathbf{x}}$ and $\hat{\mathbf{A}}\cdot\hat{\mathbf{p}}$ interaction Hamiltonian,s, using successively the dipole approximation and the exact coupling.
\end{abstract}


\maketitle

\section{Introduction \label{sec:Intro}} 

We study the spontaneous emission of light by an atomic dipole (two-level quantum system) at zero temperature in which an electronic state is initially excited while all the electromagnetic modes are taken to be in their vacuum (zero-photon) state. In the standard treatment of this kind of problem, one derives the Fermi golden rule \cite{Cohen2}, which states that the probability that the dipole stays in its excited state (the so-called survival probability) decreases linearly with time. At longer times, the standard treatment is the Wigner-Weisskopf approach \cite{WW}, in the framework of which the survival probability is predicted to decrease exponentially. It is well known \cite{MisraSudarshan}, on the other hand, that, under very general circumstances, the survival probability of a quantum system decays quadratically -and not linearly- with time at very short times. This initial quadratic behaviour is known as the quantum Zeno regime.

We studied in detail the behaviour of an atomic dipole in the Zeno regime. Our calculations predict a small but sudden decrease of the survival probability, in conjunction with the emission of off-resonant light. For the $2\mathrm{p}-1\mathrm{s}$ transition in atomic hydrogen -within the rotating wave approximation- we found that electromagnetic modes with frequency up to a thousand times larger than that of the atomic transition are excited.

As we shall show, the transient Zeno regime is dominated by the collective, in phase, response of the electromagnetic field modes. Very quickly, this collective response gets out of phase (incoherent) so that non-resonant electromagnetic modes become, in a way, inactive. It is only at this point that the linear decay of the survival probability arises, as described by the Fermi golden rule. This linear decay is essentially driven by the resonant modes of the electromagnetic field.

Our paper is organized as follows. In Sect.~\ref{sec:QDipole} we summarize the standard quantum electrodynamical (QED) description of dipolar decay, including the Fermi golden rule. In Sect.~\ref{sec:Isabelle} we describe a perturbative approach to the problem in which the outgoing electromagnetic modes are discretized. In this approach only nearly resonant (on shell) modes are considered. This approach makes it possible  to simulate an exponential decay of the survival probability in full agreement with the Wigner-Weisskopf approximation. It also predicts a Zeno effect which can be considered to be ``standard'': for short times the parabolic dependence in time of the survival probability results in a short ``plateau'' directly followed by a linear time dependence (Fermi golden rule), naturally merging at longer times into the exponential behaviour.

In Sects.~\ref{sec:EdotXDipole}, \ref{sec:EdotXExact} and \ref{sec:AdotP} we generalize the perturbative approach of Sect.~\ref{sec:Isabelle} by taking account of the response of off-resonant electromagnetic modes. For very short times, their collective response results in a non-standard Zeno effect which is the essential new result of our paper. We first study this effect in the case for the $\hat{\mathbf{E}}\cdot\hat{\mathbf{x}}$ atom-field coupling (Sects.~\ref{sec:EdotXDipole} and \ref{sec:EdotXExact}), and then for the $\hat{\mathbf{A}}\cdot\hat{\mathbf{p}}$ coupling (Sect.~\ref{sec:AdotP}). In both cases, the collective answer of the off-resonant modes induces a small decrease of the survival probability for short times, leading to the usual Fermi and Wigner-Weisskopf regimes. The last section is devoted to open questions and conclusions.

\section{Basic elements on the decay of a quantum dipole \label{sec:QDipole}} 

\subsection{Hamiltonian of the atom-light interaction \label{subsec:QED}} 
We consider a two-level atom (ground state $\mid\!\mathrm{g}\rangle$, excited state $\mid\!\mathrm{e}\rangle$) interacting with the electromagnetic field. The atom sits either in free space or in an ideal metallic cavity. In both cases, the field can be expanded in terms of normal modes labelled by an index $\lambda$. The Hamiltonian reads
\begin{equation} \label{eq:QEDHamilton}
  \hat{H}=\hat{H}_A+\hat{H}_R+\hat{H}_I,
\end{equation}
where
\begin{equation} \label{eq:AtomHamilton}
  \hat{H}_A=E_{\mathrm{g}} \ \mid\!\mathrm{g}\rangle\langle \mathrm{g}\!\mid + E_{\mathrm{e}} \ \mid\!\mathrm{e}\rangle\langle \mathrm{e}\!\mid
\end{equation}
is the atomic Hamiltonian ($E_{\mathrm{g}}$ and $E_{\mathrm{e}}$ being the energies of the ground and excited state), 
\begin{equation} \label{eq:FieldHamilton}
  \hat{H}_R=\sum_{\lambda} \hbar \omega_\lambda\left(\hat{a}_\lambda^\dagger \hat{a}_\lambda+\frac{1}{2}\right)
\end{equation}
is the Hamiltonian of the free radiation ($\hat{a}_\lambda$ is the annihilation operator for photons in the mode $\lambda$ and $\omega_\lambda$ the angular frequency of these photons). Let us consider for the moment that the atom-field interaction Hamiltonian reads
\begin{equation} \label{eq:IntHamilton}
  \hat{H}_I=-\hat{\mathbf{d}}\cdot\frac{\hat{\mathbf{D}}\left(\mathbf{r}_0\right)}{\epsilon_0},
\end{equation}
which is valid in the electric dipole approximation. Here, $\hat{\mathbf{d}}$ is the atomic dipole operator and $\hat{\mathbf{D}}\left(\mathbf{r}_0\right)$ is the displacement field at the atom location $\mathbf{r}_0$. We shall go beyond this approximation later but for the time being we prefer to formulate the problem at this level of approximation, having in mind that the techniques that we shall develop in the next sections can easily be generalized to other choices of interaction Hamiltonians. Within the two-level atom approximation, the dipole $\mathbf{d}$ can be written as
\begin{equation} \label{eq:DipoleMatrix}
  \hat{\mathbf{d}}=\hat{\idmx} \ \hat{\mathbf{d}} \ \hat{\idmx} = \sum_{i,j}\mid\! i\rangle\langle i\!\mid \hat{\mathbf{d}}\mid\! j\rangle \langle j\!\mid= \sum_{i,j} \mathbf{d}_{ij} \ \mid\! i\rangle\langle j\!\mid,
\end{equation}
where $\left(i,j\right)\in\{\mathrm{e},\mathrm{g}\}^2$ and  $\mathbf{d}_{ij}=\langle i\!\mid \hat{\mathbf{d}} \mid\! j \rangle$. Because of spherical symmetry, $\mathbf{d}_{\mathrm{gg}}=\mathbf{d}_{\mathrm{ee}}=0$ and $\mathbf{d}_{\mathrm{ge}}=\mathbf{d}_{\mathrm{eg}}^*$. Therefore the electric dipole operator can be written
\begin{equation} \label{eq:Dipole}
  \hat{\mathbf{d}}=\mathbf{d}_{\mathrm{ge}} \mid\!\mathrm{g}\rangle\langle \mathrm{e}\!\mid +\mathbf{d}_{\mathrm{ge}}^* \mid\!\mathrm{e}\rangle\langle \mathrm{g}\!\mid. 
\end{equation}
Since there are no free charges in the system, the displacement field is purely transverse everywhere in space (in contrast to the electric field which is not transverse at $\mathbf{r}_0$) and can be expanded on the transverse modal functions $\mathbf{v}_\lambda^T$ as
\begin{equation}\label{eq:ElectricField}
  \frac{\hat{\mathbf{D}}\left(\mathbf{r}\right)}{\epsilon_0}= \mathrm{i} \sum_\lambda \sqrt{\frac{\hbar \omega_\lambda}{2\epsilon_0}}\left( \hat{a}_\lambda \ \mathbf{v}_\lambda^T\left(\mathbf{r}\right) -  \hat{a}_\lambda^\dagger \ \mathbf{v}_\lambda^{T*}\left(\mathbf{r}\right)\right)
\end{equation}
if the modal functions are normalized such that 
\begin{equation} \label{eq:NormIntegral}
  \int \mathbf{v}_{\lambda}^T\left(\mathbf{r}\right)\cdot \mathbf{v}_{\lambda'}^{T*}\left(\mathbf{r}\right) \ \mathrm{d}\mathbf{r}= \delta_{\lambda \lambda'}.
\end{equation}
By plugging the expansions (\ref{eq:Dipole}) and (\ref{eq:ElectricField}) into (\ref{eq:IntHamilton}), we can write the interaction Hamiltonian $\hat{H}_I$ as the sum of two contributions $\hat{H}_I^{R}$ and $\hat{H}_I^{AR}$, where 
\begin{equation} \label{eq:Resonant}
  \hat{H}_I^R=\mathrm{i} \hbar \sum_\lambda \left(G_{\lambda}\left(\mathbf{r}_0\right) \ \hat{a}_\lambda^\dagger \ \mid\!\mathrm{g}\rangle\langle \mathrm{e}\!\mid - G_{\lambda}^*\left(\mathbf{r}_0\right) \ \hat{a}_\lambda \ \mid\!\mathrm{e}\rangle\langle \mathrm{g}\!\mid\right)
\end{equation}
contains the \emph{rotating} interaction terms and
\begin{equation} \label{eq:AntiResonant}
  \hat{H}_I^{AR}=\mathrm{i} \hbar \sum_\lambda \left(F_{\lambda}\left(\mathbf{r}_0\right) \ \hat{a}_\lambda^\dagger \ \mid\!\mathrm{e}\rangle\langle \mathrm{g}\!\mid - F_{\lambda}^*\left(\mathbf{r}_0\right) \ \hat{a}_\lambda \ \mid\!\mathrm{g}\rangle\langle \mathrm{e}\!\mid\right)
\end{equation}
contains the \emph{counter-rotating} interaction terms. The coupling constants $G_\lambda\left(\mathbf{r}_0\right)$ and $F_\lambda\left(\mathbf{r}_0\right)$ are given by
\begin{subequations} \label{eq:Couplings}
  \begin{align}
    G_\lambda\left(\mathbf{r}_0\right) &=\sqrt{\frac{\omega_\lambda}{2 \hbar \epsilon_0}}\  \mathbf{d}_{\mathrm{ge}}\cdot \mathbf{v}_\lambda^{T*}\left(\mathbf{r}_0\right), \label{eq:GCouple}\\
    F_\lambda\left(\mathbf{r}_0\right) &=\sqrt{\frac{\omega_\lambda}{2 \hbar \epsilon_0}}\  \mathbf{d}_{\mathrm{ge}}^*\cdot \mathbf{v}_\lambda^{T*}\left(\mathbf{r}_0\right).\label{eq:FCouple}
  \end{align}
\end{subequations}
The interaction terms appearing in $\hat{H}_I^R$ (\ref{eq:Resonant}) are easy to interpret. The first term corresponds to the creation of a photon while the atom jumps from its excited to its ground state. The second term represents the reverse process, that is, photon absorption.

The interaction Hamiltonian $\hat{H}_I^{AR}$ (\ref{eq:AntiResonant}) is sometimes claimed \emph{not} to conserve energy \footnote{Of course, energy \emph{is} conserved because the QED Hamiltonian (\ref{eq:QEDHamilton}) is time-independent. From this point of view, even $\hat{H}_I^R$ apparently violates energy when photons which are off-resonant with the atomic transition are considered. This discussion is in general omitted because it is common to focus on resonant electromagnetic modes only, which is only justified in the long time limit as we shall show.}. The first term represents the process of adding one photon to the field while simultaneously promoting the atom from the ground to the excited state; the second one represents the reverse process.  In the derivation of the Fermi golden rule as well as in the Wigner-Weisskopf approach, it is common to neglect the counter-rotating interaction terms $\hat{H}_I^{AR}$, which is called the rotating-wave approximation (RWA) \cite{CohenQED2,PassanteVirtual}. As our goal is ultimately to estimate corrections to the W-W predictions, we shall from now on confine ourselves to the same regime and thus stick to the Rotating Wave Approximation. This is particularly justified in studies of the short-time behaviour of the system, because to first order in perturbation theory, the counter-rotating interaction Hamiltonian (\ref{eq:AntiResonant}) does not contribute. On the other hand, we shall not neglect the emission of off-resonant photons as is traditionally done.

\subsection{Decay equations within the Rotating Wave Approximation  \label{subsec:DecayEq}}

We now consider the problem of an atom that is initially in its excited state $\mid\!\mathrm{e}\rangle$ and decays to its ground state $\mid\!\mathrm{g}\rangle$ due to its coupling to the electromagnetic field. If the field is initially unpopulated, the quantum state of the system at any time can be written as
\begin{equation}\label{eq:Ansatz}
  \begin{split}
    \mid\!\psi\left(t\right)\rangle &= c_{\mathrm{e}}\left(t\right) \ \mathrm{e}^{-\frac{\mathrm{i}}{\hbar} \left(E_{\mathrm{e}}+E_{\mathrm{vac}}\right) t} \ \mid\!\mathrm{e},0\rangle\\
    &+ \sum_\lambda c_{\mathrm{g},\lambda}\left(t\right) \ \mathrm{e}^{-\frac{\mathrm{i}}{\hbar} \left(E_{\mathrm{g}}+\hbar \omega_\lambda+E_{\mathrm{vac}}\right) t} \ \mid\!\mathrm{g},1_\lambda\rangle,
  \end{split}
\end{equation}
where $\mid\!\mathrm{e},0\rangle$ means that the atom is in the excited state and the field contains no photons and $\mid\!\mathrm{g},1_\lambda\rangle$ means that the atom is in the ground state and the field contains a single photon in the mode $\lambda$. In the absence of any interaction ($\hat{H}_I=0$), the coefficients $c_{\mathrm{e}}$ and $c_{\mathrm{g},\lambda}$ do not evolve and each term of the superposition oscillates at its own eigenfrequency. In the so-called weak coupling regime, the functions $c_{\mathrm{e}}$ and $c_{\mathrm{g},\lambda}$ are expected to be slowly varying in time. The vacuum energy $E_{\mathrm{vac}}=\sum_\lambda\hbar\omega_\lambda/2$ produces a phase shift oscillating at an infinite angular frequency $E_{\mathrm{vac}}/\hbar$. However, this phase shift can be factored out and appears as a global phase that can be simply ignored.

By injecting (\ref{eq:Ansatz}) in the Schrödinger equation
\begin{equation}
  \mathrm{i} \hbar \frac{\mathrm{d}}{\mathrm{d} t} \mid\!\psi\left(t\right)\rangle= \left(\hat{H}_A+\hat{H}_R+\hat{H}_I^R\right) \mid\!\psi\left(t\right)\rangle, 
\end{equation} 
we find the differential equations governing the evolution of functions $c_{\mathrm{e}}$ and $c_{\mathrm{g},\lambda}$. They read
\begin{subequations} \label{eq:ExactEQ}
  \begin{align}
    \dot{c}_{\mathrm{e}}\left(t\right) &= - \sum_{\lambda} G_\lambda^*\left(\mathbf{r}_0\right) \ c_{\mathrm{g},\lambda}\left(t\right) \ \mathrm{e}^{-\mathrm{i} \left(\omega_\lambda-\omega_{\mathrm{eg}}\right) t}, \label{eq:ExactEQe}\\
    \dot{c}_{\mathrm{g},\lambda}\left(t\right) &=   G_\lambda\left(\mathbf{r}_0\right) \ c_{\mathrm{e}}\left(t\right) \ \mathrm{e}^{\mathrm{i} \left(\omega_\lambda-\omega_{\mathrm{eg}}\right)  t}.\label{eq:ExactEQg}
  \end{align}
\end{subequations}
Up to the RWA, these equations are exact and apply to both emission in free-space and cavity emission. They possess no known exact solution \cite{Cohen2}.

Wigner and Weisskopf \cite{WW} attempted to solve these equations. They analysed the case of spontaneous emission in \emph{free space} using the ansatz 
\begin{equation} \label{eq:WWAnsatz}
  c_{\mathrm{e}}\left(t\right)\approx \mathrm{e}^{-\frac{\Gamma}{2}t}
\end{equation}
to explain the universally observed exponential decay of the excited state population. This solution is not an exact solution, but is widely considered to be a satisfactory approximation at long times. Using this ansatz, Wigner and Weisskopf found that $\Gamma$ is well approximated by
\begin{equation} \label{eq:GammaRate}
  \Gamma=\frac{\omega_{\mathrm{eg}}^3 \left|\mathbf{d}_{\mathrm{ge}}\right|^2}{3\pi\hbar\epsilon_0 c^3},
\end{equation}
which coincides with the value found by applying the Fermi golden rule, as we now discuss.

\subsection{Fermi golden rule from time-independent perturbation theory \label{subsec:NoTimeTime}}

Consider the interaction Hamiltonian $\hat{H}_I$ as a perturbation. The Fermi golden rule is usually obtained from time-dependent perturbation theory but here we show how time-independent perturbation theory to the first order also allows to retrieve the Fermi golden rule. This original approach opens the way to a more general time-independent perturbative treatment of the problem that will be developed later. The eigenstates of the free Hamiltonian $\hat{H}_A+\hat{H}_R$ are $\mid\!\mathrm{e},0\rangle$ and $\mid\!\mathrm{g},1_\lambda\rangle$, with respective eigenenergies $E_{\mathrm{e}}=\hbar\omega_{\mathrm{e}}$ and $E_{\mathrm{g}}+\hbar\omega_\lambda=\hbar\left(\omega_{\mathrm{g}}+\omega_\lambda\right)$. Perturbation theory \cite{Cohen2} tells us that, treating $\hat{H}_I$ to first order, these eigenenergies are unmodified. The normalized perturbed eigenstates are
\begin{subequations} \label{eq:PertStates}
  \begin{align}
    \mid\!\mathrm{e},0\rangle^{\left(1\right)}&=\frac{1}{\sqrt{N}}\left(\mid\!\mathrm{e},0\rangle+\mathrm{i}\sum_\lambda\frac{G_\lambda\left(\mathbf{r}_0\right)}{\omega_{\mathrm{eg}}-\omega_\lambda}\mid\!\mathrm{g},1_\lambda\rangle\right), \label{eq:PrimeExcited}\\
    \mid\!\mathrm{g},1_\lambda\rangle^{\left(1\right)}&=\frac{1}{\sqrt{N_\lambda}}\left(\mid\!\mathrm{g},1_\lambda\rangle+\mathrm{i}\frac{G_\lambda^*\left(\mathbf{r}_0\right)}{\omega_{\mathrm{eg}}-\omega_\lambda}\mid\!\mathrm{e},0\rangle\right) \label{eq:PrimeGround}
  \end{align}
\end{subequations}
where
\begin{subequations} \label{eq:NormGalore}
  \begin{align}
    N&=1+\sum_\lambda\frac{\left|G_\lambda\left(\mathbf{r}_0\right)\right|^2}{\left(\omega_{\mathrm{eg}}-\omega_\lambda\right)^2}, \label{eq:BigNorm}\\
    N_\lambda&=1+\frac{\left|G_\lambda\left(\mathbf{r}_0\right)\right|^2}{\left(\omega_{\mathrm{eg}}-\omega_\lambda\right)^2}. \label{eq:NotAsBigaNorm}
  \end{align}
\end{subequations}
Note that the perturbed eigenstates (\ref{eq:PertStates}) are orthogonal to lowest order in the perturbation. Notice that, to lowest order, we have
\begin{equation} \label{eq:Thomastuce}
  \mid\!\mathrm{e},0\rangle=\frac{1}{\sqrt{N}}\left(\mid\!\mathrm{e},0\rangle^{\left(1\right)}-\mathrm{i}\sum_\lambda\sqrt{\frac{N_\lambda}{N}}\frac{G_\lambda\left(\mathbf{r}_0\right)}{\omega_{\mathrm{eg}}-\omega_\lambda}\mid\!\mathrm{g},1_\lambda\rangle^{\left(1\right)}\right).
\end{equation}
We can thus write the evolution of the system when the atom is initially excited, with no photon in the field. Using the spectral theorem for the total Hamiltonian $\hat{H}$ with its approximate eigenstates (\ref{eq:PertStates}) derived from time-independent perturbation theory, we get
\begin{equation} \label{eq:SurvivalAmpl}
\begin{split}
\langle \mathrm{e},0\!\mid\mathrm{e}^{-\frac{\mathrm{i}}{\hbar}\hat{H}t}\mid\!\mathrm{e},0\rangle=& \frac{\mathrm{e}^{-\mathrm{i}\omega_{\mathrm{eg}}t}}{N} \times \\
&\left(1+\sum_\lambda\left|G_\lambda\left(\mathbf{r}_0\right)\right|^2\frac{\mathrm{e}^{\mathrm{i}\left(\omega_{\mathrm{eg}}-\omega_\lambda\right)t}}{\left(\omega_{\mathrm{eg}}-\omega_\lambda\right)^2}\right).
\end{split}
\end{equation}
After a bit of algebra, this yields
\begin{equation} \label{eq:SurvivalSine}
\begin{split}
\left|\langle \mathrm{e},0\!\mid\mathrm{e}^{-\frac{\mathrm{i}}{\hbar}\hat{H}t}\mid\!\mathrm{e},0\rangle\right|^2=1-\frac{4}{N^2}\sum_\lambda &\frac{\left|G_\lambda\left(\mathbf{r}_0\right)\right|^2}{\left(\omega_{\mathrm{eg}}-\omega_\lambda\right)^2} \times\\
&\sin^2\left(\left(\omega_{\mathrm{eg}}-\omega_\lambda\right)\frac{t}{2}\right)
\end{split}
\end{equation}
to lowest order. Let us define $P_{\mathrm{decay}}$ as 1 minus the survival probability: $P_{\mathrm{decay}}\left(t\right)\equiv1-\left|\langle \mathrm{e},0\!\mid\mathrm{e}^{-\frac{\mathrm{i}}{\hbar}\hat{H}t}\mid\!\mathrm{e},0\rangle\right|^2$. In free space, the sum over free modes is an integral. The substitution follows the procedure
\begin{equation} \label{eq:DiscretetoContinuous}
  \frac{1}{V}\sum_\lambda\rightarrow\sum_{\varkappa=1}^2\int\frac{\mathrm{d}^3k}{\left(2\pi\right)^3}
\end{equation}
where $\varkappa$ are the two polarizations, $V$ is the initial quantization volume and the modal functions are given by
\begin{equation} \label{eq:VacuumModes}
  \mathbf{v}_\lambda^T\left(\mathbf{r}\right)\rightarrow\mathbf{v}_\varkappa^T\left(\mathbf{k},\mathbf{r}\right)=\frac{\mathrm{e}^{\mathrm{i}\mathbf{k}\cdot\mathbf{r}}}{\sqrt{V}}\bm{\epsilon}_{\left(\varkappa\right)}\left(\mathbf{k}\right).
\end{equation}
The unit polarization vectors obey \cite{Weinberg} the closure relation
\begin{equation} \label{eq:PolarSum}
  \sum_{\varkappa=1}^2\left(\epsilon_{\left(\varkappa\right)}^i\right)^*\left(\mathbf{k}\right)\epsilon_{\left(\varkappa\right)}^j\left(\mathbf{k}\right)=\delta^{ij}-\frac{k^ik^j}{\mathbf{k}^2}.
\end{equation}
With all this in mind we compute the decay probability
\begin{equation} \label{eq:DecayFights}
\begin{split}
    P_{\mathrm{decay}}\left(t\right)=&\frac{4Vc}{2\hbar\epsilon_0}\sum_{\varkappa=1}^2\int\frac{\mathrm{d}^3k}{\left(2\pi\right)^3}\frac{\left|\mathbf{d}_{\mathrm{ge}}\cdot\bm{\epsilon}_{\left(\varkappa\right)}^*\left(\mathbf{k}\right)\right|^2}{V}\\
    &\times\left|\mathbf{k}\right| \ \frac{\sin^2\left(\left(\omega_{\mathrm{eg}}-c\left|\mathbf{k}\right|\right)\frac{t}{2}\right)}{\left(\omega_{\mathrm{eg}}-c\left|\mathbf{k}\right|\right)^2}\\
    =&\frac{2c}{\hbar\epsilon_0}\int\frac{\mathrm{d}^3k}{\left(2\pi\right)^3}\left[\left|\mathbf{d}_{\mathrm{ge}}\right|^2-\left|\mathbf{d}_{\mathrm{ge}}\cdot\frac{\mathbf{k}}{\left|\mathbf{k}\right|}\right|^2\right]\\
    &\times\left|\mathbf{k}\right| \ \frac{\sin^2\left(\left(\omega_{\mathrm{eg}}-c\left|\mathbf{k}\right|\right)\frac{t}{2}\right)}{\left(\omega_{\mathrm{eg}}-c\left|\mathbf{k}\right|\right)^2}\\
    =&\frac{t^2 \ \left|\mathbf{d}_{\mathrm{ge}}\right|^2}{6\pi^2\epsilon_0\hbar c^3}\int_0^{\infty}\mathrm{d}\omega \ \omega^3 \ \sin\!\mathrm{c}^2\left(\left(\omega_{\mathrm{eg}}-\omega\right)\frac{t}{2}\right)\\
\end{split}
\end{equation}
where we performed the angular integration by requiring that the polar angle $\theta$ be zero when $\mathbf{k}$ points in the direction of $\mathbf{d}_{\mathrm{ge}}$.

In the limit of large times, that is, times larger than $1/\omega_{\mathrm{eg}}$, one finds the usual Fermi golden rule with the decay rate $\Gamma$ (\ref{eq:GammaRate}), that is, $P_{\mathrm{decay}}\left(t\right)=\Gamma t$.   In this regime, the cardinal sine in (\ref{eq:DecayFights}) becomes very peaked around the atomic transition frequency. One can then consider that only frequencies close to $\omega_{\mathrm{eg}}$ contribute to (\ref{eq:DecayFights}) and rewrite the integral in the latter equation as
\begin{equation*}\int_{\omega_{\mathrm{eg}}-\Delta}^{\omega_{\mathrm{eg}}+\Delta}\mathrm{d}\omega\,\omega^3\sin\!\mathrm{c}^2\left(\left(\omega_{\mathrm{eg}}-\omega\right)\frac{t}{2}\right).
\end{equation*}
Then, making use of
\begin{equation} \label{eq:WeAreDone}
\begin{split}
&\int_{\omega_{\mathrm{eg}}-\Delta}^{\omega_{\mathrm{eg}}+\Delta} \mathrm{d}\omega \ \omega^3 \ \sin\!\mathrm{c}^2 \  \left[\left(\omega_{\mathrm{eg}}-\omega\right)\frac{t}{2}\right]\\
&\underset{t\rightarrow+\infty}{\sim}\omega_{\mathrm{eg}}^3 \ \int_{-\infty}^{+\infty} \mathrm{d}\omega \ \sin\!\mathrm{c}^2 \ \left[\left(\omega_{\mathrm{eg}}-\omega\right)\frac{t}{2}\right]\\
    &=\omega_{\mathrm{eg}}^3 \ \frac{2\pi}{t},
\end{split}
\end{equation}
the Fermi Golden Rule is established. This is only valid at so-called intermediate times, larger than $1/\omega_{\mathrm{eg}}$, but smaller than $1/\Gamma$, so that the perturbative approach still makes sense.

Strictly speaking, taking the limit $t\rightarrow+\infty$ under the integral on the right-hand side of (\ref{eq:DecayFights}), as is usually done to obtain the Fermi golden rule, is not rigorous, since that integral diverges (for all times). For instance, for sufficiently small times the cardinal sine in (\ref{eq:DecayFights}) can be taken to be equal to 1, and hence the integral behaves like  $\int_0^{+\infty}\mathrm{d}\omega\,\omega^3=+\infty$. To accommodate this, we introduce a cutoff on frequencies. A crude but natural 
\footnote{If we go beyond the dipolar approximation, the (exact) coupling factors are proportional to matrix elements of the form \unexpanded{$ \langle i\!\mid \hat{\mathbf{x}} \exp\left(\mathrm{i}\mathbf{k}\cdot\hat{\mathbf{x}}\right)\mid\! j \rangle$} 
for the $\hat{\mathbf{E}}\cdot\hat{\mathbf{x}}$ coupling and \unexpanded{$\langle i\!\mid \nabla \exp\left(\mathrm{i}\mathbf{k}\cdot\hat{\mathbf{x}}\right)\mid\! j \rangle$} for the $\hat{\mathbf{A}}\cdot\hat{\mathbf{p}}$ coupling  where $i$ and $j$ represent either $1\mathrm{s}$ or $2\mathrm{p}$ electronic states. When the wave number $\left|\left|\mathbf{k}\right|\right|$ becomes higher than the inverse of the Bohr radius, such matrix elements de facto vanish because the oscillating exponential $\exp\left(\mathrm{i}\mathbf{k}\cdot\hat{\mathbf{x}}\right)$ averages out during the integration.
} 
cutoff frequency is the one corresponding to the size of the atomic dipole we considered: $K_{\mathrm{C}}=2e\pi/\left|\mathbf{d}_{\mathrm{ge}}\right|$ with $e$ the elementary charge. Note that in a recent treatment of a similar problem \cite{Norge}, the Compton frequency was used as a cutoff. We shall see later that when one goes beyond the dipole approximation, a natural cutoff appears from the computations and need not be implemented artificially (it is of the same order of magnitude as $K_{\mathrm{C}}$). We then have, for very small times,
\begin{equation} \label{eq:OnToZeno}
  \begin{split}
    P_{\mathrm{decay}}\left(t\right)&=\frac{t^2  \left|\mathbf{d}_{\mathrm{ge}}\right|^2}{6\pi^2  \epsilon_0  \hbar   c^3} \ \int_0^{c\,K_{\mathrm{C}}}\mathrm{d}\omega\,\omega^3\\
    &=\frac{c}{24\pi^2\epsilon_0\hbar} \ \left|\mathbf{d}_{\mathrm{ge}}\right|^2 \ K_{\mathrm{C}}^4 \ t^2.
  \end{split}
\end{equation}
A parabolic Zeno behaviour is thusly obtained. To connect this initial behaviour with that, at longer times, predicted by the Fermi golden rule and the Wigner-Weisskopf approximation, we make appeal to numerical calculations. As a prerequisite, we now investigate how results from Wigner and Weisskopf are retrieved, in a discretized approach in which the two-level atom is coupled with a finite number of quasiresonant modes of the electromagnetic field.

\section{Exponential decay for a two-level atom coupled with quasiresonant field modes \label{sec:Isabelle}} 

\subsection{A discrete approach to spontaneous emission \label{subsec:Finitist}} 

Instead of a continuum of field modes, we will now consider a finite amount thereof. We will thus pass from a continuous regime where Fermi's Golden rule can be applied to a discrete regime where the sum in (\ref{eq:Ansatz}) is both discrete and finite. The Hilbert space is accordingly assumed to be finite-dimensional, which opens the way to a considerably simplified numerical treatment based on matrix equations.

The problem that we face is twofold: for each energy there is a continuous infinity of field modes which corresponds to all possible spatial directions of emission of a photon; moreover we must consider a range of energies running from 0 to $\hbar cK_{\mathrm{C}}$. Usually, only resonant (on-shell) photons are considered, and the multiplicity of outgoing photonic modes is hidden in the Density of States (DoS) formulation of the Fermi golden rule \cite{Cohen2}. We shall adopt here another strategy, in which we integrate over the continuum of spatial directions of emission (and over various polarizations) without resorting to any kind of approximation.

In order to explain the principle of our method, let us first consider a situation in which two electromagnetic modes of equal energy only are present. The total (interaction + free) and free Hamiltonians then read
\begin{multline} \label{eq:MatrixForm}
  H= \begin{bmatrix}
    \hbar\omega_0 & -\mathrm{i}\hbar G_1^* & -\mathrm{i}\hbar G_2^* \\
    \mathrm{i}\hbar G_1 & \hbar\omega & 0 \\
    \mathrm{i}\hbar G_2 & 0 & \hbar\omega
  \end{bmatrix}
  \hfill
  H_0= \begin{bmatrix}
    \hbar\omega_0 & 0 & 0 \\
    0 & \hbar\omega & 0 \\
    0 & 0 & \hbar\omega
  \end{bmatrix}.
\end{multline}
The corresponding Schrödinger equation is  
\begin{multline}
  \mathrm{i}\hbar\frac{\mathrm{d}}{\mathrm{d}t}
  \begin{bmatrix}
    x_0\left(t\right) \\
    x_1\left(t\right) \\
    x_2\left(t\right)
  \end{bmatrix}
  =
  H_0
  \begin{bmatrix}
    x_0\left(t\right) \\
    x_1\left(t\right) \\
    x_2\left(t\right)
  \end{bmatrix}
  \\+\mathrm{i}\hbar
  \begin{bmatrix}
    - G_1^*x_1\left(t\right)- G_2^*x_2\left(t\right) \\
    G_1x_0\left(t\right) \\
    G_2x_0\left(t\right)
  \end{bmatrix}.
\end{multline}
Let us now define the states $|A\rangle$ and $|B\rangle$ through
\begin{align}
  |A\rangle&={\mathrm{i}\over \sqrt{|G_1|^2+|G_2|^2}} \
  \begin{bmatrix}
    0 \\
    G_1 \\
    G_2
  \end{bmatrix},
  \\
  |B\rangle&={\mathrm{i}\over \sqrt{|G_1|^2+|G_2|^2}} \ 
  \begin{bmatrix}
    0 \\
    G_2^* \\
    -G_1^*
  \end{bmatrix}.
\end{align}
In the new orthonormal basis spanned by the states $(1 \ 0 \ 0)^T$, $|A\rangle$ and $|B\rangle$, the Schrödinger equation now reads

\begin{equation} \label{eq:Decoupled}
  \mathrm{i}\frac{\mathrm{d}}{\mathrm{d}t}
  \begin{bmatrix}
    x_0\left(t\right) \\
    x_A\left(t\right) \\
    x_B\left(t\right)
  \end{bmatrix}
  =
  \begin{bmatrix}
     G_{\mathrm{eff}}\,x_A\left(t\right)+\omega_0 \,x_0\left(t\right) \\
     G_{\mathrm{eff}}\,x_0\left(t\right)+\omega \,x_A\left(t\right) \\
    \omega \,x_B\left(t\right)
  \end{bmatrix},
\end{equation}
where we introduced the effective coupling constant $G_{\mathrm{eff}}=\sqrt{|G_1|^2+|G_2|^2}$.
Obviously, the interaction only involves the state $|A\rangle$, while the orthogonal state $|B\rangle$ ignores the coupling with the ``excited atom'' state $(1 \ 0 \ 0)^T$.
This is still true even if an arbitrary number of modes (say $N$) are coupled to the excited state: the states orthogonal to 
\begin{equation} \label{eq:EffectiveState}
  |A\rangle={1\over \sqrt{|G_1|^2+|G_2|^2+\cdots+|G_N|^2}}
  \begin{bmatrix}
    0 \\
    G_1 \\
    G_2 \\
    \cdots\\
    G_N
  \end{bmatrix}
\end{equation}
are not coupled to the excited state in the case where we consider modes with a same energy $\omega$. We are thus free to replace the whole set of modes by an effective state $|A\rangle$ characterized by an effective coupling constant which obeys 
\begin{equation} \label{eq:EffectiveCoupling}
  G_{\mathrm{eff}}=\sqrt{|G_1|^2+|G_2|^2+\cdots+|G_N|^2},
\end{equation}
that can be considered equivalent to the summation rule of the $\Gamma$ factors over various decay channels which is traditionally used in the context of the Fermi golden rule in QED 
and particle physics as well.
However, in order to numerically integrate over slices of various energies, we have to discretize the energy, in order to implement an approximation scheme that we can tackle numerically.

Making use of equations (\ref{eq:Resonant}), (\ref{eq:Couplings}), (\ref{eq:DiscretetoContinuous}) and (\ref{eq:EffectiveCoupling}), it is easy to check that a slice of energy comprised between $\hbar\,c\,k_\lambda$ and $\hbar\,c\,k_{\lambda+1}$ is characterized by an effective coupling constant equal to 
\begin{equation} \label{eq:ThisTripleIntegral}
\begin{split}
G_{\mathrm{eff}}^{\left(\lambda\right)}= \left[\sum_{\varkappa=1}^2\int_0^{2\pi}\mathrm{d}\varphi\right. &  \int_0^\pi\mathrm{d}\theta 
 \ \sin\theta  \times \\
&\left. \int_{k_\lambda}^{k_{\lambda+1}} \frac{k^2 \ \mathrm{d}k}{\left(2\pi\right)^3} \left|G_\varkappa\left(\mathbf{k},\mathbf{r}_0\right) \right|^2\right]^{\frac{1}{2}}.
\end{split}
\end{equation}
As shown in App.~\ref{subsubsec:EdotXDipoleAppdx} where the effective Hamiltonian is rewritten in terms of (averaged) ladder operators, if we successively perform the integration over the angles and over energies, we find
\begin{equation}G_{\mathrm{eff}}^{\left(\lambda\right)}=\frac{1}{2}\sqrt{\frac{\hbar c}{6\epsilon_0}}\left|\mathbf{d}_{\mathrm{ge}}\right|\left(k_{\lambda+1}^4-k_\lambda^4\right)^{\frac{1}{2}}.
\label{eq:HIntShell}
\end{equation}
We also fixed (see App.~\ref{subsubsec:EdotXDipoleAppdx}) the energy of the effective mode assigned to an energy slice by imposing that it be equal to the average energy over the slice, weighted by the DoS. This yields, for an energy slice bounded by $k_\lambda$ and $k_{\lambda+1}$, the effective energy $\omega_{\mathrm{eff}}^{\left(\lambda\right)}=\left(4/5\right)\hbar c\left(k_{\lambda+1}^5-k_\lambda^5\right)/\left(k_{\lambda+1}^4-k_\lambda^4\right)$. From now on, we drop the $_{\mathrm{eff}}$ subscripts from these quantities. The Hamiltonian of the system is as before of the form
\begin{equation} \label{eq:IsaHamiltonian}
  \begin{aligned} [b]
    \hat{H}&=\hbar\omega_{\mathrm{eg}}\vert\mathrm{e},0\rangle\langle\mathrm{e},0\vert +\sum_\lambda \hbar\omega_\lambda \vert \mathrm{g},1_\lambda\rangle\langle \mathrm{g},1_\lambda\vert\\
    &+\mathrm{i}\hbar\sum_\lambda \left[ G_\lambda\left(\mathbf{r}_0\right) \vert \mathrm{g},1_\lambda\rangle\langle\mathrm{e},0\vert - G_\lambda^*\left(\mathbf{r}_0\right) \vert\mathrm{e},0\rangle\langle \mathrm{g},1_\lambda\vert \right],
  \end{aligned}
\end{equation}
excepted that here again summations run over a finite amount of modes. We describe the dynamics of this system by solving the Schrödinger equation
\begin{equation} \label{eq:IsaSchroedinger}
  \hat{H}\vert\psi\left(t\right)\rangle=\mathrm{i}\hbar\frac{\mathrm{d}}{\mathrm{d}t}\vert\psi\left(t\right)\rangle.
\end{equation}
which can be set in matrix form as
\begin{multline} \label{eq:MatrixFormAgain}
  \frac{\mathrm{d}}{\mathrm{d}t}
  \begin{bmatrix}
    c_{\mathrm{e}}\left(t\right) \\
    c_{\mathrm{g},1}\left(t\right) \\
    \vdots \\
    c_{\mathrm{g},N_{\mathrm{comb}}}\left(t\right)
  \end{bmatrix}\\
  = \begin{bmatrix}
    -\mathrm{i}\omega_{\mathrm{eg}} & - G^*\left(\omega_1\right) & \ldots & - G^*\left(\omega_{N_{\mathrm{comb}}}\right) \\
     G\left(\omega_1\right) & -\mathrm{i}\omega_1 & & \\
    \vdots & & \ddots & \text{\huge0}\\
     G \left(\omega_{N_{\mathrm{comb}}}\right) & \text{\huge0}& & -\mathrm{i}\omega_{N_{\mathrm{comb}}}
  \end{bmatrix} \begin{bmatrix}
    c_{\mathrm{e}}\left(t\right) \\
    c_{\mathrm{g},1}\left(t\right) \\
    \vdots \\
    c_{\mathrm{g},N_{\mathrm{comb}}}\left(t\right)
  \end{bmatrix}.
\end{multline}
The eigenvalues ($\equiv \kappa_\lambda$) and eigenvectors ($\equiv \vert \kappa_\lambda \rangle$) of $\hat{H}$ are estimated numerically and thereafter are used to solve for $c_{\mathrm{e}}\left(t\right)$ and $c_{\mathrm{g},\lambda}\left(t\right)$:
\begin{equation}
  \begin{bmatrix}
    c_{\mathrm{e}}\left(t\right) \\
    c_{\mathrm{g},1}\left(t\right) \\
    \vdots \\
    c_{\mathrm{g},N_{\mathrm{comb}}}\left(t\right)
  \end{bmatrix} = \sum_{\lambda=1}^{N+1}\alpha_j \vert \kappa_j\rangle \mathrm{e}^{-\mathrm{i}\kappa_j t}.
\end{equation}
To describe a situation in which the atom is initially excited, in a zero temperature environment, we consider the initial conditions
\begin{equation}
  \left\{ \begin{array}{lcl}
    c_{\mathrm{e}}\left(t=0\right)&=&1,\\
    c_{\mathrm{g},\lambda}\left(t=0\right)&=&0
  \end{array} \right.
  \label{eq:incond}
\end{equation}

\subsection{Numerical results \label{subsec:Results}} 
We now consider $N_{\mathrm{comb}}$ electromagnetic modes forming a frequency comb around $\omega_{\mathrm{eg}}$. The frequency spacing between two successive modes is taken to be much smaller than $\omega_{\mathrm{eg}}$. If one wants to generate the dynamics of the Fermi regime, one must make sure that the total frequency range $\Delta\omega$ of these electromagnetic field modes satisfies
\begin{equation} \label{eq:ToGetFermi}
  \Delta\omega\gtrapprox \Gamma
\end{equation}
with $\Gamma$ given by (\ref{eq:GammaRate}), which can be seen as a manifestation of the time-energy uncertainty $\Delta E\Delta t \geqslant \hbar/2$.

If $\Delta\omega$ is taken to be much smaller than advised by (\ref{eq:ToGetFermi}) then some modes which are significantly excited by their interaction with the atom are wrongly ignored. If $\Delta\omega$ is taken to be much larger, then off-resonant modes are needlessly included in the treatment \cite{Isabelle}.
\begin{figure}[t]
  \begin{center}
    \includegraphics[width=1\columnwidth]{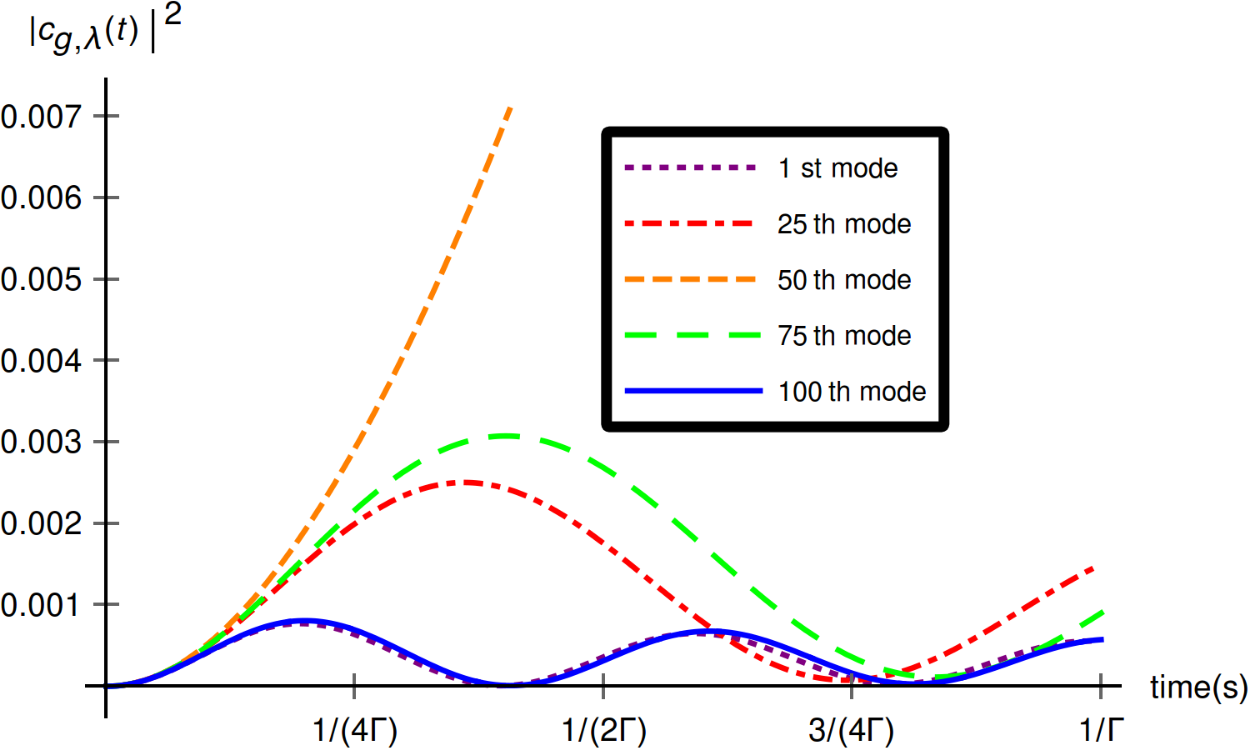}
  \end{center}
  \caption{Plot of $\left|c_{\mathrm{g},\lambda}\left(t\right)\right|^2$ for various modes for a system with $N_{\mathrm{comb}}=100$ resonant modes with frequencies within $\Delta\omega=16 \ \Gamma$ of $\omega_{\mathrm{eg}}$. Note that at short times the behaviour is quadratic for all modes: this reflects the initial, coherent response of the electromagnetic vacuum.}
  \label{fig:ActiveSmall}
\end{figure}
One can indeed distinguish the ``active" from the ``inactive" modes by looking at the $\left|c_{\mathrm{g},\lambda}\left(t\right)\right|^2$ curves of Fig.~\ref{fig:ActiveSmall}.

\begin{figure}[tp!]
  \begin{center}
    \includegraphics[width=1\columnwidth]{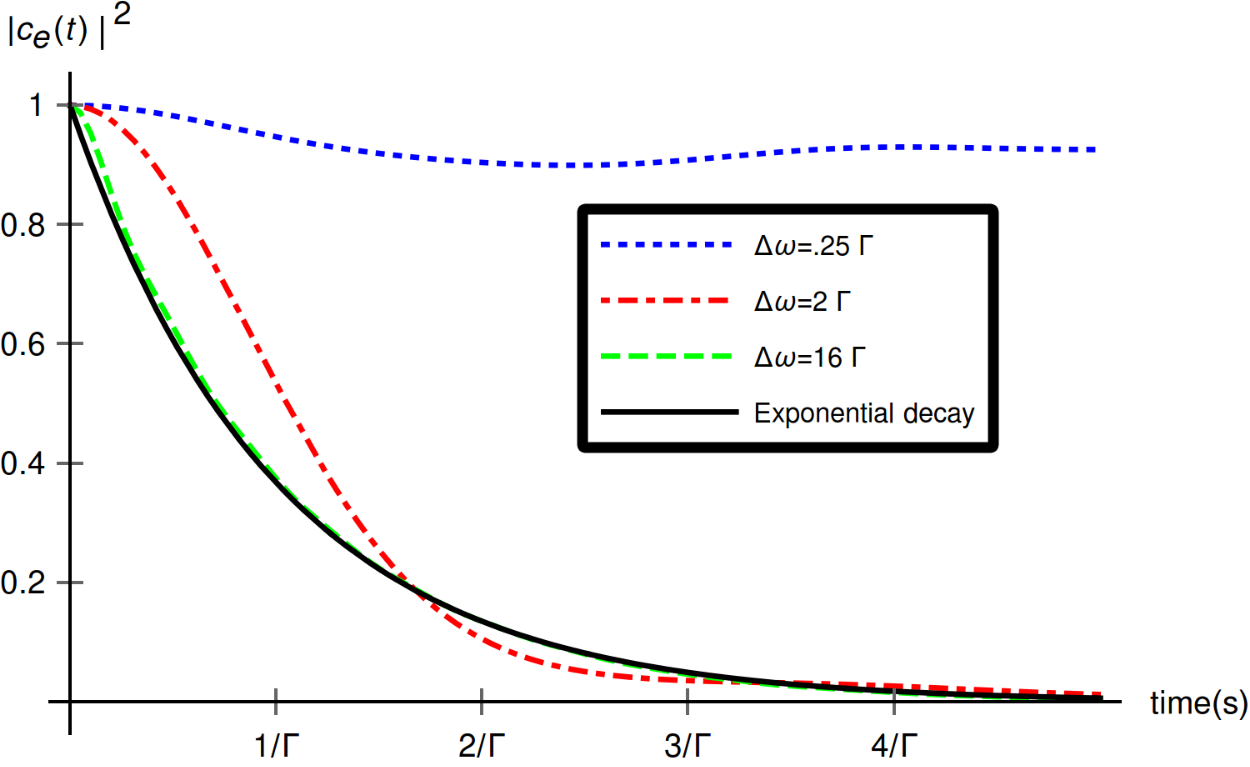}
  \end{center}
  \caption{Transition towards exponential decay by fulfilling better and better the condition $\Delta\omega\gtrapprox\Gamma$.}
  \label{fig:AlphaFermi}
\end{figure}

In practice, we varied the comb width $\Delta\omega$ in order to look for the change in behaviour when $\Delta\omega$ approached a value close to $\Gamma$. As to the probability amplitudes $\left|c_{\mathrm{g},\lambda}\left(t\right)\right|^2$ of the field modes, one would expect from the expression for $\left|c_{\mathrm{g},\lambda}\left(t\right)\right|^2$ (see (\ref{eq:DecayFights})) to see a quadratic curve at short times. Notice that this initial behaviour should be entirely independent of frequency. These expectations are nicely met in Fig.~\ref{fig:ActiveSmall}. Fig.~\ref{fig:AlphaFermi} shows how one gets closer to the Fermi regime -which constitutes the beginning of the exponential decay- by widening the electromagnetic frequency comb.\\

\begin{figure}[t]
  \begin{center}
    \includegraphics[width=1\columnwidth]{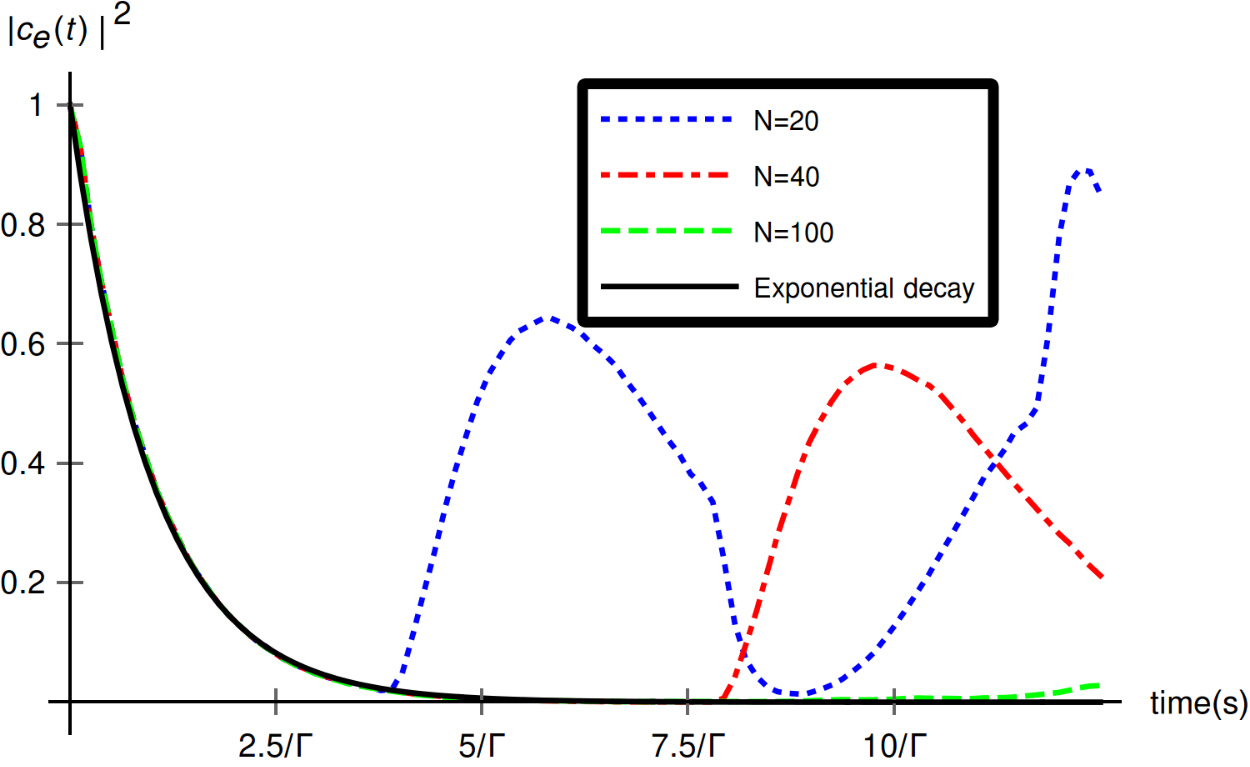}
  \end{center}
  \caption{Decay of the survival probability $\vert c_{\mathrm{e}}\left(t\right)\vert^2$ for calculations with $N_{\mathrm{comb}}=20$ (dotted blue), $N_{\mathrm{comb}}=40$ (dot-dashed red) and $N_{\mathrm{comb}}=100$ (dashed green) against the theoretical expected behaviour $\exp\left(-\Gamma t\right)$ (solid black). The curves for $N=20$, $N=40$ and $N=100$ are virtually indistinguishable for $t<2.5/\Gamma$ (remember $1/\Gamma=\SI{1.60d-9}{s}$).}
  \label{fig:AlphaFit}
\end{figure}

We used, for our numerical calculations, the experimental values for the $2 \text{p}- 1 \text{s}$ transition in atomic hydrogen, namely, $\omega_{\mathrm{eg}}=\SI{1.55d16}{s}^{-1}$, $\left|\mathbf{d}_{\mathrm{ge}}\right|=\SI{6.31d-30}{C m}$ and $1/\Gamma=\SI{1.60d-9}{s}$  \cite{BetheSalpeter}. The mode discretization described above and in the App. is quite robust in terms of its dependence on $N_{\mathrm{comb}}$, as illustrated by Fig.~\ref{fig:AlphaFit}, where we compared the survival probability $\vert c_{\mathrm{e}}\left(t\right)\vert^2$ found from numerical calculations to the standard result of Wigner-Weisskopf perturbation theory.

The survival probability displays a ``revival'' due to the discretization of the outgoing modes, which results in a finite recurrence time $t_{\mathrm{recur}}$. However, a number of $N_{\mathrm{comb}}=100$ quasiresonant modes is enough to send this recurrence time to around $10/\Gamma$ (see Fig.~\ref{fig:AlphaFit}), which is more than enough for what we want to study here. Even with $N_{\mathrm{comb}}=20,$ the discrepancy with the W-W theory in the region of interest (say $0<t<2.5 / \Gamma\approx\SI{4d-8}{s}$) is not noticeable.
 Note that the condition for the ``weak coupling limit", $\Gamma \ll \omega_{\mathrm{eg}}$, is de facto met for the system we considered. We can also detect on Fig.~\ref{fig:AlphaFermi} (see curve with $\Delta\omega=16 \ \Gamma$) the initial Zeno regime after which the curve indeed approximates the Wigner-Weisskopf exponential. We shall consider later the role played by high-frequency, off-resonant modes in the Zeno zone.

\begin{figure}[t]
  \begin{center}
    \includegraphics[width=1\columnwidth]{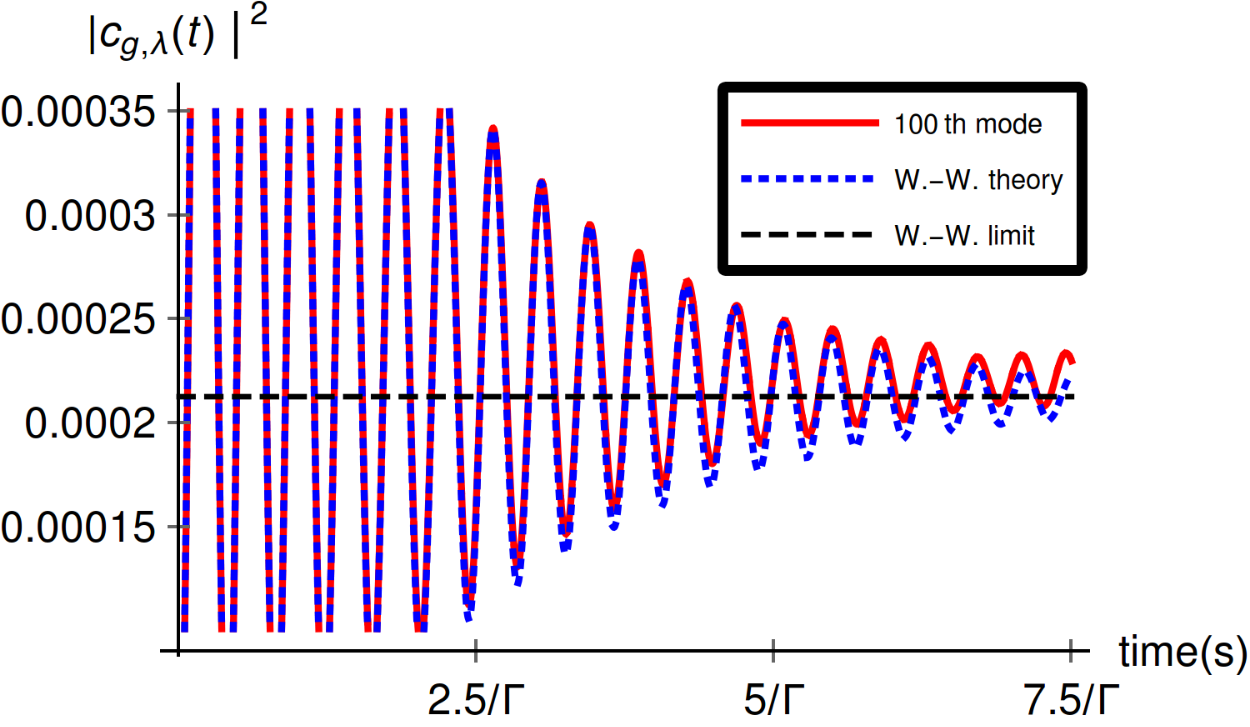}
  \end{center}
  \caption{Evolution, for a system with $N_{\mathrm{comb}}=100$ quasiresonant modes of the field, of the probability that the $100^{\text{th}}$ mode of the field is occupied (solid red), plotted against the expected Wigner-Weisskopf behaviour (\ref{eq:ModeProba}) (dotted blue) and the asymptotic limit (\ref{eq:SatVal}) of the latter expression (dashed black). \label{fig:Fairly}}
\end{figure}

\begin{figure}[t]
  \begin{center}
    \includegraphics[width=1\columnwidth]{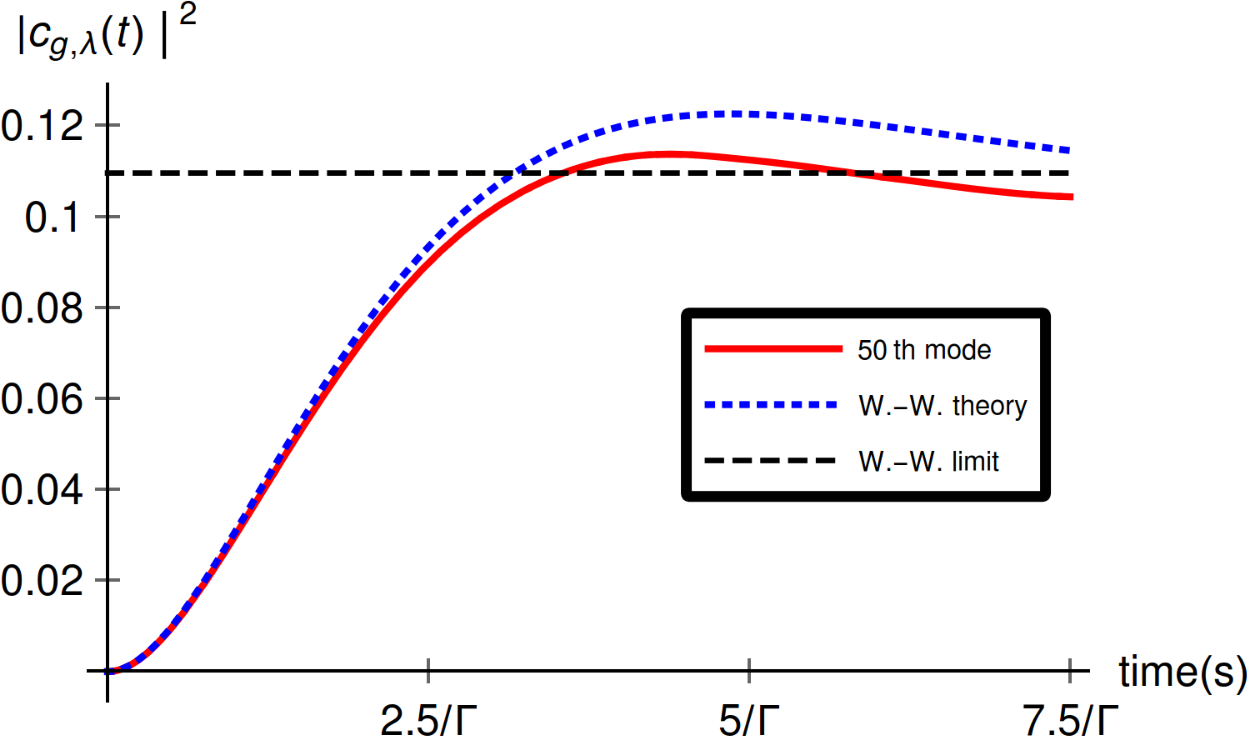}
  \end{center}
  \caption{Evolution, for a system with $N_{\mathrm{comb}}=100$ quasiresonant modes of the field, of the probability that the (very resonant) $50^{\text{th}}$ mode of the field is occupied (solid red), plotted against the expected Wigner-Weisskopf behaviour (\ref{eq:ModeProba}) (dotted blue) and the asymptotic limit (\ref{eq:SatVal}) of the latter expression (dashed black). \label{fig:Very}}
\end{figure}

Our results agree with the Fermi golden rule in the interval $\omega_{\mathrm{eg}}^{-1}<t<\Gamma$ but also with the Wigner-Weisskopf perturbation theory for longer times. For $N_{\mathrm{comb}}=100$ quasiresonant modes, we plotted in Figs.~\ref{fig:Fairly} and \ref{fig:Very} the time evolution of $\vert c_{\mathrm{g},\lambda}\left(t\right)\vert^2$ for two modes, one rather close to the resonant transition frequency (Fig.~\ref{fig:Fairly}), and the other one extremely close to resonance (Fig.~\ref{fig:Very}). Wigner-Weisskopf theory predicts \cite{Englert}
\begin{equation} \label{eq:ModeProba}
  c_{\mathrm{g},\lambda}\left(t\right)=\mathrm{e}^{-\mathrm{i}\omega_\lambda t}\,G_\lambda\left(\mathbf{r}_0\right)\,\frac{\mathrm{e}^{-\mathrm{i}\left(\omega_{\mathrm{eg}}-\omega_\lambda\right)t}\mathrm{e}^{-\Gamma t/2}-1}{\omega_\lambda-\omega_{\mathrm{eg}}+\mathrm{i}\Gamma/2} ,
\end{equation}
which tends to
\begin{equation} \label{eq:SatVal}
  \vert c_{\mathrm{g},\lambda}\left(t\right)\vert^2 \stackrel{\Gamma t\rightarrow + \infty}{\longrightarrow} \frac{\vert G_\lambda\left(\mathbf{r}_0\right)\vert^2}{(\omega_\lambda-\omega_{\mathrm{eg}})^2+(\Gamma/2)^2}
\end{equation}
as $t\rightarrow+\infty$. These expectations are nicely met too in Figs.~\ref{fig:Fairly} and \ref{fig:Very} where the asymptotic limits (\ref{eq:SatVal}) are explicitly drawn.\\

We investigate in several steps the influence of off-resonant electromagnetic modes on the decay of the survival probability.

First (Sect.~\ref{sec:EdotXDipole}), we use the $\hat{\mathbf{E}}\cdot\hat{\mathbf{x}}$ interaction Hamiltonian in the dipole approximation of Sect.~\ref{subsec:QED} and discard all frequencies above the cutoff frequency $c K_{\mathrm{C}}$ of Sect.~\ref{subsec:NoTimeTime}, while in Sect.~\ref{sec:EdotXExact} we make use of the exact matrix elements for the $\hat{\mathbf{E}}\cdot\hat{\mathbf{x}}$ interaction Hamiltonian.

Then (Sect.~\ref{sec:AdotP}), we repeat the same process with the $\hat{\mathbf{A}}\cdot\hat{\mathbf{p}}$ interaction Hamiltonian. In Sect.~\ref{subsec:AdotPDipole} we use the dipole-approximated coupling derived similarly to that of Sect.~\ref{subsec:QED} and discard all frequencies above the cutoff frequency $c K_{\mathrm{C}}$ of Sect.~\ref{subsec:NoTimeTime}, while in Sect.~\ref{subsec:AdotPExact} we make use of the exact matrix elements for the $\hat{\mathbf{A}}\cdot\hat{\mathbf{p}}$ interaction Hamiltonian found in \cite{FacchiPhD}.

\section{$\hat{\mathbf{E}}\cdot\hat{\mathbf{x}}$ coupling in the dipole approximation \label{sec:EdotXDipole}}

\begin{figure}[t]
  \begin{center}
    \includegraphics[width=1\columnwidth]{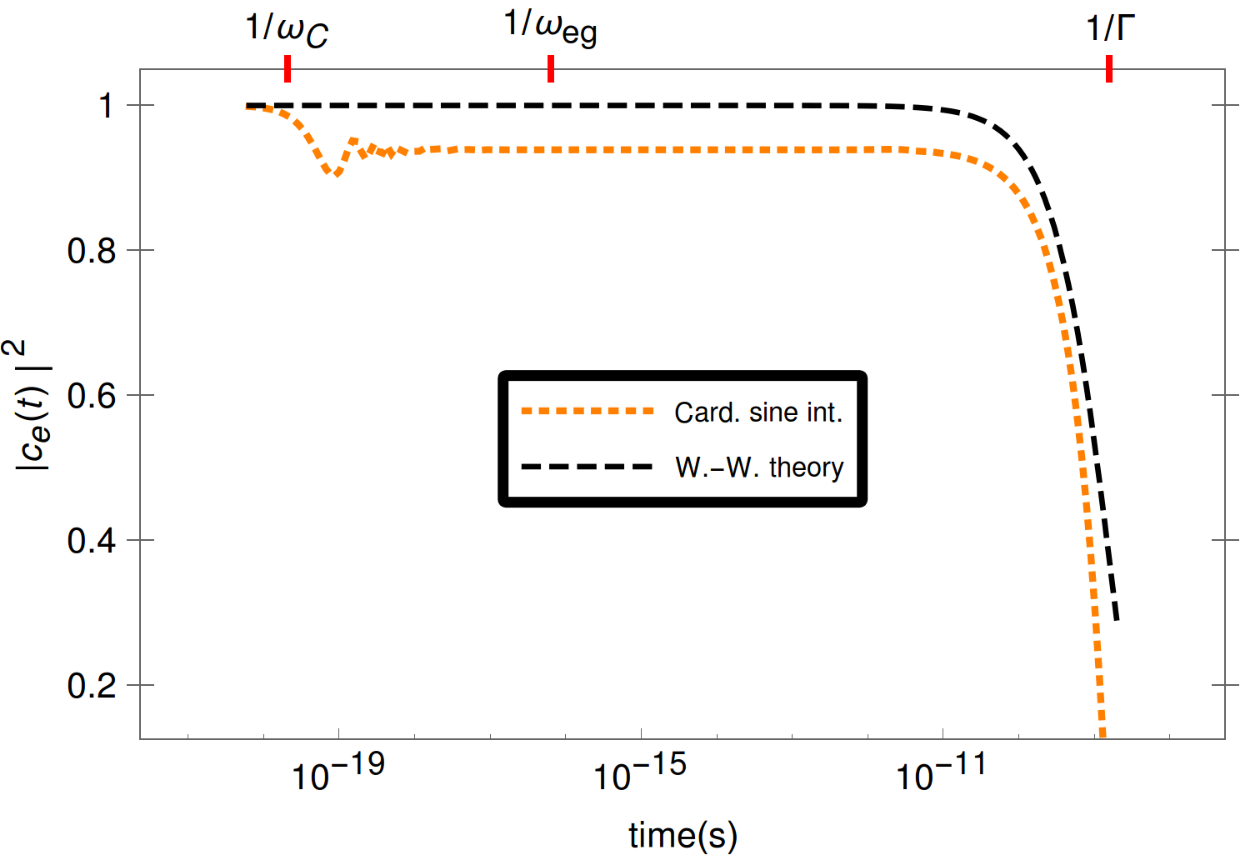}
  \end{center}
  \caption{Linear-log plot of the survival probability $\left|c_{\mathrm{e}}\left(t\right)\right|^2$ from  Wigner-Weisskopf theory (dashed black) and the truncated cardinal sine integral (\ref{eq:CardSine}) (dotted orange) with $\left|c_{\mathrm{e}}\left(t\right)\right|^2=1-P_{\mathrm{decay}}\left(t\right)$. The two main time regimes (strong emission at very short times and exponential decay later) can be visualized on this single graphic. Remember $1/\omega_{\mathrm{C}}=\SI{2.09d-20}{s}$, $1/\omega_{\mathrm{eg}}=\SI{6.45d-17}{s}$,  $1/\Gamma=\SI{1.60d-9}{s}$.}
  \label{fig:WWVSSC}
\end{figure}

We focus here on the dipole approximation. Remember that the integral (\ref{eq:DecayFights}) diverges at all times, which requires us to introduce a cutoff frequency, as was done in Sect.~\ref{subsec:NoTimeTime}. If we follow this prescription, the integral (\ref{eq:DecayFights}) becomes
\begin{equation} \label{eq:CardSine}
  P_{\mathrm{decay}}\left(t\right)=\frac{t^2  \left|\mathbf{d}_{\mathrm{ge}}\right|^2}{6\pi^2  \epsilon_0  \hbar   c^3} \int_0^{c\,K_{\mathrm{C}}}\mathrm{d}\omega\,\omega^3\sin\!\mathrm{c}^2\left(\left(\omega_{\mathrm{eg}}-\omega\right)\frac{t}{2}\right).
\end{equation}
This expression of the decay probability is compared with the result of the usual Wigner-Weisskopf approximation on Fig.~\ref{fig:WWVSSC}. One notices, at very short times -the so-called Zeno regime- a strong drop during the quadratic decay of the truncated cardinal sine integral (\ref{eq:CardSine}). At longer times, the integral (\ref{eq:CardSine}) follows the linear decay predicted by Fermi's golden rule -with an offset- and fails to reproduce the predictions of the Wigner-Weisskopf approximation beyond $t>1/\Gamma$, which is expected from this first-order approach. As we shall show this discrepancy can be suppressed by going beyond first order perturbation theory to describe quasiresonant electromagnetic modes.\\

Our goal here is to connect the two regimes (Zeno for short times and W-W for long times) observed on Fig.~\ref{fig:WWVSSC}. For intermediate times, where the Fermi golden rule is valid, we know that first order perturbation prevails. Besides, it is notoriously known that perturbation theory is not accurate when we consider non-perturbed states with very close energies. In order to develop a perturbative approach, we treated the quasiresonant modes first by diagonalising the Hamiltonian as done in Sect.~\ref{sec:Isabelle}, before incorporating the off-resonant modes of the electromagnetic field through time-independent perturbation theory. Let us denote by $\hat{H}_0$ the Hamiltonian which describes the atom, the field, and the interaction between the atom and the quasiresonant modes of the electromagnetic field, and $\hat{H}_{\mathrm{high}}$ the Hamiltonian which describes the interaction between the atom and the other (off-resonant) modes. Namely
\begin{subequations} \label{eq:CombThenShell}
\begin{equation} \label{eq:MessiahHamiltonian}
\begin{aligned} [b]
      \hat{H}_0&=\hbar\omega_{\mathrm{eg}}\vert\mathrm{e},0\rangle\langle\mathrm{e},0\vert + \sum_{\lambda=1}^{N_{\mathrm{comb}}+N_{\mathrm{offshell}}}\hbar\omega_\lambda \vert \mathrm{g},1_\lambda\rangle\langle \mathrm{g},1_\lambda\vert\\
      &+\mathrm{i}\hbar\sum_{\lambda=1}^{N_{\mathrm{comb}}}\left[ G_\lambda\left(\mathbf{r}_0\right) \vert \mathrm{g},1_\lambda\rangle\langle\mathrm{e},0\vert - G_\lambda^*\left(\mathbf{r}_0\right) \vert\mathrm{e},0\rangle\langle \mathrm{g},1_\lambda\vert \right]
\end{aligned}
\end{equation}
  and
\begin{equation} \label{eq:ShellHamiltonian}
\begin{split}
    \hat{H}_{\mathrm{high}}=\mathrm{i}\hbar\sum_{\lambda=N_{\mathrm{comb}}+1}^{N_{\mathrm{comb}}+N_{\mathrm{offshell}}}
    & \left[ G_\lambda\left(\mathbf{r}_0\right) \vert \mathrm{g},1_\lambda\rangle\langle\mathrm{e},0\vert \right. \\
     & \left. - G_\lambda^*\left(\mathbf{r}_0\right) \vert\mathrm{e},0\rangle\langle \mathrm{g},1_\lambda\vert \right]
\end{split}
  \end{equation}
\end{subequations}
where $N_{\mathrm{comb}}$ is, as previously, the number of quasiresonant modes and $N_{\mathrm{offshell}}$ is the number of off-resonant modes. The effective frequencies $\omega_\lambda$ and couplings $G_\lambda\left(\mathbf{r}_0\right)$ are defined in App.~\ref{subsubsec:EdotXDipoleAppdx}. We wrote $\vert\mathrm{g},1_\lambda\rangle=\vert\mathrm{g}\rangle\otimes \vert1_\lambda\rangle$ where $\vert\mathrm{g}\rangle$ is the atomic ground state while the single photon-single mode states $\vert1_\lambda\rangle$ are obtained by the action of the creation operators $\left(\hat{a}_{\mathrm{avg}}^\lambda\right)^\dagger$ (\ref{eq:AverageA}) on the electromagnetic vacuum $\vert0\rangle$. Now, assume that we diagonalized $\hat{H}_0$. Since the off-resonant modes are uncoupled, the spectral decomposition of $\hat{H}_0$ reads
\begin{subequations} \label{eq:H0Diag}
  \begin{equation} \label{eq:SpectralTh}
    \hat{H}_0=\sum_{i=0}^{N_{\mathrm{comb}}}\hbar\omega_{\mathrm{eg}}^i\mid\!\psi_i\rangle\langle \psi_i\!\mid+\sum_{\lambda=N_{\mathrm{comb}}+1}^{N_{\mathrm{comb}}+N_{\mathrm{offshell}}}\hbar\omega_\lambda\mid\!\mathrm{g},1_\lambda\rangle\langle\mathrm{g},1_\lambda\!\mid
  \end{equation}
  with
  \begin{equation} \label{eq:ProjMessiah}
    \mid\!\psi_i\rangle=\epsilon^i_0\mid\!\mathrm{e},0\rangle+\sum_{\lambda=1}^{N_{\mathrm{comb}}}\epsilon^i_\lambda\mid\!\mathrm{g},1_\lambda\rangle\equiv\sum_{\lambda=0}^{N_{\mathrm{comb}}}\epsilon^i_\lambda\mid\!\lambda\rangle,
  \end{equation}
\end{subequations}
where we introduced the new notations $|\mathrm{e},0\rangle=|0\rangle$, and $|\mathrm{g},1_\lambda\rangle$=$|\lambda\rangle$. It is useful to notice that (\ref{eq:ProjMessiah}) can be inverted by
\begin{equation} \label{eq:NotObv}
  |\lambda\rangle=\sum_{\lambda=0}^{N_{\mathrm{comb}}}\left(\epsilon_\lambda^j\right)^*|\psi_j\rangle.
\end{equation}
The $\epsilon^i_\lambda$ coefficients in (\ref{eq:ProjMessiah}) were obtained numerically in the previous Sect.~\ref{sec:Isabelle} by diagonalising the matrix (\ref{eq:MatrixForm}). Expanding the relations $\langle\lambda\!\mid\!\zeta\rangle=\delta_{\lambda\zeta}$ and $\langle\psi_k\!\mid\!\psi_j\rangle=\delta_{jk}$ yields, respectively, the useful relations
\begin{subequations} \label{eq:EpsilonGalore}
  \begin{align}
    \sum_{i=0}^{N_{\mathrm{comb}}}\left(\epsilon^i_\lambda\right)^*\epsilon^i_\zeta&=\delta_{\lambda\zeta},\\
    \sum_{\lambda=0}^{N_{\mathrm{comb}}}\left(\epsilon^k_i\right)^*\epsilon^j_\lambda&=\delta_{jk}.
  \end{align}
\end{subequations}
Treating $\hat{H}_{\mathrm{high}}$ to first order in perturbation we get the normalized perturbed eigenstates
\begin{subequations} \label{eq:PertState}
  \begin{align}
    |\psi_i\rangle^{\left(1\right)}&=\frac{1}{\sqrt{N_i}}\left[|\psi_i\rangle+\mathrm{i}\epsilon^i_0\sum_{\lambda=N_{\mathrm{comb}}+1}^{N_{\mathrm{comb}}+
    N_{\mathrm{offshell}}}\frac{G_\lambda\left(\mathbf{r}_0\right)}{\left(\omega_0^i-\omega_\lambda\right)}|\mathrm{g},1_\lambda\rangle\right],\\
    |\mathrm{g},1_\lambda\rangle^{\left(1\right)}&=\frac{1}{\sqrt{N_\lambda}}\left[|\mathrm{g},1_\lambda\rangle+\mathrm{i}\frac{G_\lambda^*\left(\mathbf{r}_0\right)}{\left(\omega_{\mathrm{eg}}-\omega_\lambda\right)}|\mathrm{e},0\rangle\right]
  \end{align}
\end{subequations}
with
\begin{subequations} \label{eq:NormGaloreAgain}
  \begin{align}
    N_i&=1+\left|\epsilon^i_0\right|^2\sum_{\lambda=N_{\mathrm{comb}}}^{N_{\mathrm{comb}}+N_{\mathrm{offshell}}}\frac{\left|G_\lambda\left(\mathbf{r}_0\right)\right|^2}{\left(\omega_0^i-\omega_\lambda\right)^2}, \label{eq:MediumNorm}\\
    N_\lambda&=1+\frac{\left|G_\lambda\left(\mathbf{r}_0\right)\right|^2}{\left(\omega_{\mathrm{eg}}-\omega_\lambda\right)^2}. \label{eq:LazyNorm}
  \end{align}
\end{subequations}
Note that to first order in $G_\lambda\left(\mathbf{r}_0\right)/\left[\left(\omega_{\mathrm{eg}}-\omega_\lambda\right)\right]$, these perturbed states are orthogonal. This means that, still to first order in these parameters, the perturbed spectral decomposition of $\hat{H}=\hat{H}_0+\hat{H}_{\mathrm{high}}$ and the corresponding evolution operator are valid.

\begin{figure}[t]
  \begin{center}
    \includegraphics[width=1\columnwidth]{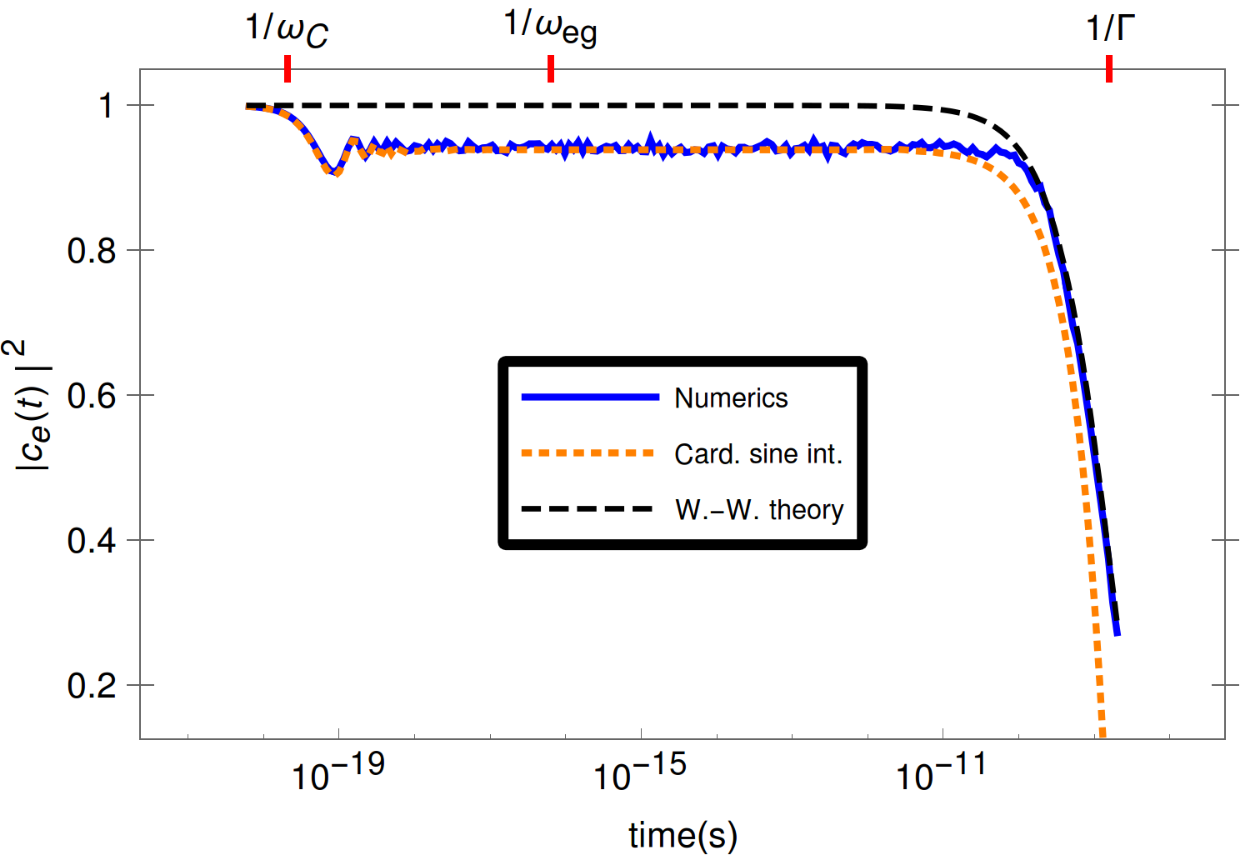}
  \end{center}
  \caption{Decay of the survival probability $\left|c_{\mathrm{e}}\left(t\right)\right|^2$ with the dipole-approximated $\hat{\mathbf{E}}\cdot\hat{\mathbf{x}}$ coupling from numerical calculations (solid blue), Wigner-Weisskopf theory (dashed black) and the truncated cardinal sine integral (\ref{eq:CardSine}) (dotted orange) with $\left|c_{\mathrm{e}}\left(t\right)\right|^2=1-P_{\mathrm{decay}}\left(t\right)$. The logarithmic scale for time is used so that the two main time regimes (strong emission at very short times and exponential decay later), as well as the transition between them, can be visualized on this single graphic. Quasiresonant modes are discretized with $N_{\mathrm{comb}}=100$ and off-resonant modes are discretized with $N_{\mathrm{offshell}}=100$. Remember $1/\omega_{\mathrm{C}}=\SI{2.09d-20}{s}$, $1/\omega_{\mathrm{eg}}=\SI{6.45d-17}{s}$ and $1/\Gamma=\SI{1.60d-9}{s}$. \label{fig:Final}}
\end{figure}

The initially excited state $\mid\!\varphi\left(t=0\right)\rangle=\,\mid\!\mathrm{e},0\rangle$ can be expanded over the perturbed eigenstates, making use of (\ref{eq:EpsilonGalore}) and (\ref{eq:PertState}):
\begin{equation} \label{eq:PerturbExpand}
\begin{aligned} [b]
    |\varphi\left(t=0\right)\rangle&=\frac{1}{\sqrt{N}}\left[\sum_{\lambda=0}^{N_{\mathrm{comb}}}\sqrt{N_i}\left(\epsilon^i_0\right)^*|\psi_i\rangle^{\left(1\right)}\right.\\
      &\left.-\sum_{\lambda=N_{\mathrm{comb}}+1}^{N_{\mathrm{comb}}+N_{\mathrm{offshell}}}
      \sqrt{N_\lambda}\frac{G_\lambda\left(\mathbf{r}_0\right)}{\left(\omega_{\mathrm{eg}}-\omega_\lambda\right)}|\mathrm{g},1_\lambda\rangle^{\left(1\right)}\right]\\
    &=\frac{1}{\sqrt{N}}\left[1+\sum_{\lambda=N_{\mathrm{comb}}+1}^{N_{\mathrm{comb}}+N_{\mathrm{offshell}}}\frac{\left|G_\lambda\left(\mathbf{r}_0\right)\right|^2}{\left(\omega_{\mathrm{eg}}-\omega_\lambda\right)^2}\right]|\mathrm{e},0\rangle
  \end{aligned}
\end{equation}
and thus
\begin{equation} \label{eq:ThisNorm}
  \sqrt{N}=1+\sum_{\lambda=N_{\mathrm{comb}}+1}^{N_{\mathrm{comb}}+N_{\mathrm{offshell}}}\frac{\left|G_\lambda\left(\mathbf{r}_0\right)\right|^2}{\left(\omega_{\mathrm{eg}}-\omega_\lambda\right)^2}.
\end{equation}
Then we can compute the survival probability at time $t$. For that we let the evolution operator $\hat{U}\left(t\right)$ act on $\mid\!\varphi\left(t=0\right)\rangle$, which yields
\begin{multline} \label{eq:TimeExpand}
  \mid\!\varphi\left(t\right)\rangle=\frac{1}{\sqrt{N}}\left[\sum_{i=0}^{N_{\mathrm{comb}}}\sqrt{N_i}\left(\epsilon^i_0\right)^*\mathrm{e}^{-\mathrm{i}\omega_0^it}\mid\!\psi_i\rangle^{\left(1\right)}\right.\\
    \left.-\mathrm{i}\sum_{\lambda=N_{\mathrm{comb}}+1}^{N_{\mathrm{comb}}+N_{\mathrm{offshell}}}\sqrt{N_\lambda}\frac{G_\lambda\left(\mathbf{r}_0\right)}{\left(\omega_{\mathrm{eg}}-\omega_\lambda\right)}\mathrm{e}^{-\mathrm{i}\omega_\lambda t}\mid\!\mathrm{g},1_\lambda\rangle^{\left(1\right)}\right].
\end{multline}
The survival probability is then deduced with the help of (\ref{eq:PertState}):
\begin{equation} \label{eq:Survival}
  \begin{aligned} [b]
    P_{\mathrm{surv}}\left(t\right)&=\left|\langle\mathrm{e},0\!\mid\!\varphi\left(t\right)\rangle\right|^2\\
    &=\frac{1}{N}\left|\sum_{i=0}^{N_{\mathrm{comb}}}\left|\epsilon^i_0\right|^2\mathrm{e}^{-\mathrm{i}\omega_0^it}\right.\\
    &\hspace{37.5pt}\left.+\sum_{\lambda={N_{\mathrm{comb}}+1}}^{N_{\mathrm{comb}}+N_{\mathrm{offshell}}}\frac{\left|G_\lambda\left(\mathbf{r}_0\right)\right|^2}{\left(\omega_{\mathrm{eg}}-\omega_\lambda\right)^2}\mathrm{e}^{-\mathrm{i}\omega_\lambda t}\right|^2.
  \end{aligned}
\end{equation}
To compute this quantity we resort to numerical calculations. The main goal here is to link the initial behaviour of the survival probability as given by the cardinal sine integral (\ref{eq:DecayFights}) to the experimentally observed exponential decay at large times. The expression (\ref{eq:Survival}) of the survival probability is compared in Fig.~\ref{fig:Final} with the usual Wigner-Weisskopf exponential $\exp\left(-\Gamma t\right)$ and the truncated cardinal sine integral (\ref{eq:CardSine}).

We see on Fig.~\ref{fig:Final} that our method allows the numerical results to espouse, at short times, the predictions of first-order perturbation theory, and, at large times, the usual Wigner-Weisskopf exponential. As we shall show in the next Sect.~\ref{sec:EdotXExact}, however, the fast drop at short times seen in Figs.~\ref{fig:WWVSSC} and \ref{fig:Final} is not physical.

\section{Exact $\hat{\mathbf{E}}\cdot\hat{\mathbf{x}}$ coupling \label{sec:EdotXExact}} 

\subsection{Results \label{subsec:EdotXResults}} 

\begin{figure}[t]
  \begin{center}
    \includegraphics[width=1\columnwidth]{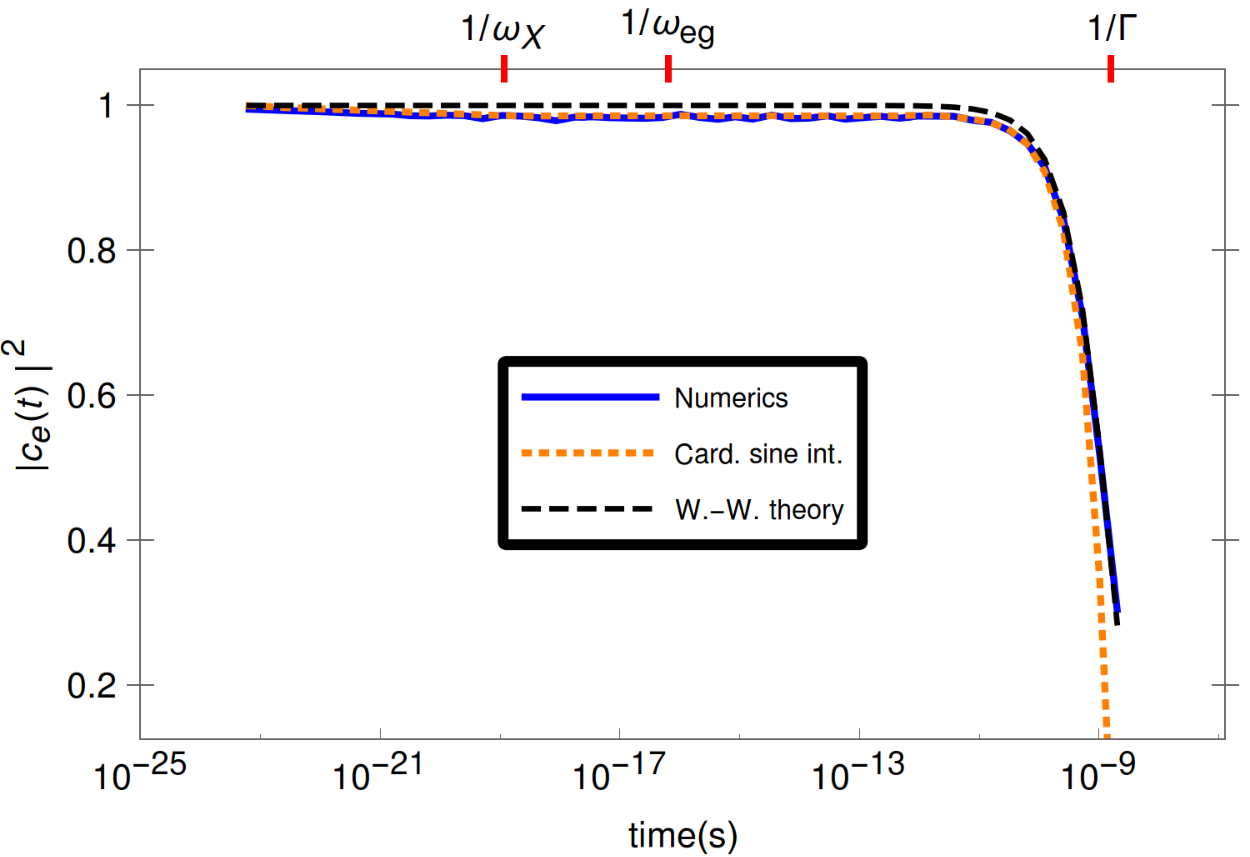}
  \end{center}
  \caption{Decay of the survival probability $\left|c_{\mathrm{e}}\left(t\right)\right|^2$ with the exact $\hat{\mathbf{E}}\cdot\hat{\mathbf{x}}$ coupling from numerical calculations (solid blue), Wigner-Weisskopf theory (dashed black) and the cardinal sine integral (\ref{eq:SeriesDecay}) (dotted orange) with $\left|c_{\mathrm{e}}\left(t\right)\right|^2=1-P_{\mathrm{decay}}\left(t\right)$. At long times numerical calculations are indistinguishable from Wigner-Weisskopf decay. A logarithmic scale for time is used. Quasiresonant modes are discretized with $N_{\mathrm{comb}}=100$ and off-resonant modes are discretized with $N_{\mathrm{offshell}}=100$. Remember $1/\omega_{\mathrm{X}}=\SI{1.18d-19}{s}$, $1/\omega_{\mathrm{eg}}=\SI{6.45d-17}{s}$ and $1/\Gamma=\SI{1.60d-9}{s}$. \label{fig:EX}}
\end{figure}

When going beyond the dipole approximation, one can derive a more exact expression for the matrix element:
\begin{multline} \label{eq:SeriesCoupling}
  G_\lambda\left(\mathbf{r}_0\right)=\frac{1}{4}\sqrt{\frac{\omega_\lambda}{2 \hbar \epsilon_0}}\  \mathbf{d}_{\mathrm{ge}}\cdot \mathbf{v}_\lambda^{T*}\left(\mathbf{r}_0\right)\\
\left[\frac{1}{\left[1+\left(\frac{\omega_\lambda}{\omega_{\mathrm{X}}}\right)^2\right]^2}+\frac{3}{2}\left(\frac{1}{1+\left(\frac{\omega_\lambda}{\omega_{\mathrm{X}}}\right)^2}+\frac{\arctan\left(\frac{\omega_\lambda}{\omega_{\mathrm{X}}}\right)}{\frac{\omega_\lambda}{\omega_{\mathrm{X}}}}\right)\right]
\end{multline}
with natural cutoff frequency $\omega_{\mathrm{X}}=\left(3c\right)/\left(2a_0\right)$, a result which, as far as we know, is original. This is done in the next Sect.~\ref{subsec:Series}. This time, first-order perturbation theory yields the following cardinal sine integral
\begin{equation} \label{eq:SeriesDecay}
\begin{split}
  P_{\mathrm{decay}}\left(t\right)=\frac{t^2 \left|\mathbf{d}_{\mathrm{ge}}\right|^2}{96\pi^2\epsilon_0\hbar c^3}  \int_0^{+\infty}\mathrm{d}\omega \ \omega^3 \ \sin\!\mathrm{c}^2\left(\left(\omega_{\mathrm{eg}}-\omega\right)\frac{t}{2}\right) \\
\left[\frac{1}{\left[1+\left(\frac{\omega}{\omega_{\mathrm{X}}}\right)^2\right]^2}+\frac{3}{2}\left(\frac{1}{1+\left(\frac{\omega}{\omega_{\mathrm{X}}}\right)^2}+\frac{\arctan\left(\frac{\omega}{\omega_{\mathrm{X}}}\right)}{\frac{\omega}{\omega_{\mathrm{X}}}}\right)\right]^2.
\end{split}
\end{equation}
The same perturbation method as that exposed in the previous Sect.~\ref{sec:EdotXDipole} is used. The frequencies $\omega_\lambda$ and couplings $G_\lambda\left(\mathbf{r}_0\right)$ are defined in App.~\ref{subsubsec:EdotXExactAppdx}. Results with this coupling are shown in Fig.~\ref{fig:EX}.

As noted above, the fast drop in the survival probability is an artifact of the truncated dipole approximation. Note that our perturbation method again gives precise results at both short, intermediate, and long times (see Fig.~\ref{fig:EX}). 

\subsection{Derivation of the interaction matrix element \label{subsec:Series}} 

We now prove (\ref{eq:SeriesCoupling}). In the Power-Zienau-Woolley picture of quantum electrodynamics, the interaction Hamiltonian reads \cite{CohenQED1}
\begin{equation} \label{eq:PZW}
\hat{H}_I=\frac{1}{\epsilon_0}\int\mathrm{d}\mathbf{x}\,\hat{\mathbf{P}}\left(\mathbf{x}\right)\cdot\hat{\mathbf{\Pi}}\left(\mathbf{x}\right).
\end{equation}
The electric polarization field reads, in the case of atomic hydrogen
\begin{equation} \label{eq:ElPol}
\hat{\mathbf{P}}\left(\mathbf{x}\right)=-e\hat{\mathbf{r}}\int_0^1\mathrm{d}u\,\delta\left(\mathbf{x}-u\hat{\mathbf{r}}\right)
\end{equation}
where $\hat{\mathbf{r}}$ is the electron position operator and $e$ the elementary electric charge (charge of a proton), and the canonically conjugate momentum to the vector potential is given by
\begin{equation} \label{eq:CanonPi}
\begin{split}
\hat{\mathbf{\Pi}}\left(\mathbf{x}\right)=\mathrm{i} \sum_{\varkappa=1}^2\int\frac{\mathrm{d}^3k}{\left(2\pi\right)^3} \sqrt{\frac{\hbar c \left|\mathbf{k}\right|}{2\epsilon_0}}\left( \hat{a}_{\left(\varkappa\right)}\left(\mathbf{k}\right)  \mathrm{e}^{\mathrm{i}\mathbf{k}\cdot\mathbf{r}}\bm{\epsilon}_{\left(\varkappa\right)}\left(\mathbf{k}\right) \right. \\
\left.-  \mathrm{h.c.}\right).
\end{split}
\end{equation}
Finally the interaction Hamiltonian can be cast in the following form:
\begin{multline} \label{eq:SeriesHere}
\hat{H}_I=\mathrm{i} \ e \ \sqrt{\frac{\hbar}{2\epsilon_0 c}} \ \hat{\mathbf{r}}\cdot\sum_{\varkappa=1}^2\int\frac{\mathrm{d}^3k}{\left(2\pi\right)^3}\sqrt{\left|\mathbf{k}\right|}\\\left[\hat{a}_{\left(\varkappa\right)}\left(\mathbf{k}\right)\bm{\epsilon}_{\left(\varkappa\right)}\left(\mathbf{k}\right)\sum_{n=0}^{+\infty}\frac{\left(\mathrm{i}\mathbf{k}\cdot\hat{\mathbf{r}}\right)^n}{\left(n+1\right)!}-\mathrm{h.c.}\right].
\end{multline}
To compute the matrix element of this interaction Hamiltonian between the $1\mathrm{s}$ and $2\mathrm{p}$ levels, the hard part is to compute $\langle1\mathrm{s}\!\mid\!\left(\mathrm{i}\mathbf{k}\cdot\hat{\mathbf{r}}\right)^n\hat{\mathbf{r}}\!\mid\!2\mathrm{p}\,m_2\rangle$ where $m_2$ is the magnetic quantum number for the $2\mathrm{p}$ sublevel considered. We follow the first steps given in \cite{FacchiPhD} for the derivation of the corresponding interaction matrix element in the $\hat{\mathbf{A}}\cdot\hat{\mathbf{p}}$ coupling.\\

Remember the expressions for the wave functions of the $1\mathrm{s}$ and $2\mathrm{p}\,m_2$ sublevels:
\begin{subequations} \label{eq:HLevels}
\begin{align}
\psi_{1\mathrm{s}}\left(\mathbf{x}\right)&=\frac{\exp\left(-\frac{\left|\mathbf{x}\right|}{a_0}\right)}{\sqrt{\pi a_0^3}}, \label{eq:Level1}\\
\psi_{2\mathrm{p}\,m_2}\left(\mathbf{x}\right)&=\frac{\exp\left(-\frac{\left|\mathbf{x}\right|}{2a_0}\right)}{8\sqrt{\pi a_0^3}}\frac{\sqrt{2}}{a_0}\mathbf{x}\cdot\bm{\xi}_{m_{2}}. \label{eq:Level2}
\end{align}
\end{subequations}
The $\bm{\xi}_{m_{2}}$ are given by
\begin{subequations} \label{eq:Xi}
\begin{align}
\bm{\xi}_0&=\mathbf{e}_z,\\
\bm{\xi}_{\pm1}&=\mp\frac{\mathbf{e}_x\pm\mathrm{i}\mathbf{e}_y}{\sqrt{2}}.
\end{align}
\end{subequations}
The three states (\ref{eq:Level2}) span the $2\mathrm{p}$ subspace for the electron, such that any state on the $2\mathrm{p}$ sublevel is a linear superposition of these three states. Choosing a coordinate system for which $\mathbf{k}/\left|\mathbf{k}\right|$ is the third basis vector, we compute
\begin{widetext}
\begin{equation} \label{eq:IntheHeart}
\begin{split}
\langle 1\mathrm{s}|\left(\mathrm{i}\mathbf{k}\cdot\hat{\mathbf{r}}\right)^n\hat{\mathbf{r}} | 2\mathrm{p}\,m_2\rangle&=\int\mathrm{d}\mathbf{x}\left(\mathrm{i}\mathbf{k}\cdot\mathbf{x}\right)^n\mathbf{x}\,\psi_{1\mathrm{s}}^*\left(\mathbf{x}\right)\psi_{2\mathrm{p}\,m_2}\left(\mathbf{x}\right)\\
&=\frac{\sqrt{2}}{8\pi}\frac{\left(\mathrm{i}\left|\mathbf{k}\right|\right)^n}{a_0^4}\int_0^{2\pi}\mathrm{d}\varphi  \int_0^\pi\mathrm{d}\theta \ \sin\theta \ \cos^n\theta \int_0^{+\infty}\mathrm{d}x \ \exp\left(-\frac{3}{2}\frac{x}{a_0}\right)x^n \ \left(\mathbf{x}\cdot\bm{\xi}_{m_{2}}\right) \ \mathbf{x}.
\end{split}
\end{equation}
In our basis where $\mathbf{k}/\left|\mathbf{k}\right|$ is the third basis vector, we have
\begin{equation} \label{eq:DotProduct}
\left(\mathbf{x}\cdot\bm{\xi}_{m_{2}}\right) \ \mathbf{x}=\left(\xi_{m_{2}}^{\left(1\right)}\sin\theta\cos\varphi
+\xi_{m_{2}}^{\left(2\right)}\sin\theta\sin\varphi+\xi_{m_{2}}^{\left(3\right)}
\cos\theta\right)
\ |\mathbf{x}|^2
\left[
\begin{array}{c}
\sin\theta\cos\varphi\\
\sin\theta\sin\varphi\\
\cos\theta
\end{array}
\right].
\end{equation}
The $\xi_{m_{2}}^{\left(i\right)}$ stand for the components of $\bm{\xi}_{m_{2}}$ in that basis. After carrying out the easy integrations over $\varphi$ and changing variables ($\eta\equiv\cos\theta$), we get
\begin{equation} \label{eq:IntheSpade}
\langle1\mathrm{s}\!\mid\!\left(\mathrm{i}\mathbf{k}\cdot\hat{\mathbf{r}}\right)^n\hat{\mathbf{r}}\!\mid\!2\mathrm{p}\,m_2\rangle =\frac{\left(\mathrm{i}\left|\mathbf{k}\right|\right)^n}{4\sqrt{2}a_0^4}\int_{-1}^1\mathrm{d}\eta
\left[
\begin{array}{c}
\xi_{m_{2}}^{\left(1\right)} \ \left(1-\eta^2\right)\\
\xi_{m_{2}}^{\left(2\right)} \ \left(1-\eta^2\right)\\
2\xi_{m_{2}}^{\left(3\right)} \ \eta^2
\end{array}
\right] \ \eta^n \int_0^{+\infty} \mathrm{d}x\,x^{4+n}\exp\left(-\frac{3}{2}\frac{x}{a_0}\right)
\end{equation}
Since, according to (\ref{eq:SeriesHere}), the dot-product of (\ref{eq:IntheSpade}) with the polarization vectors $\bm{\epsilon}_{\left(\varkappa\right)}\left(\mathbf{k}\right)$ is to be taken, its component along $\mathbf{k}/\left|\mathbf{k}\right|$ can be discarded \footnote{Remember that the polarization vectors $\bm{\epsilon}_{\left(\varkappa\right)}\left(\mathbf{k}\right)$ are orthogonal to $\mathbf{k}$, along which we chose the third axis of our basis to point.}. One can thus simply compute the scalar quantity
\begin{equation} \label{eq:OnlyTrans}
\begin{split}
\langle1\mathrm{s}|\left(\mathrm{i}\mathbf{k}\cdot\hat{\mathbf{r}}\right)^n\hat{\mathbf{r}}|2\mathrm{p}\,m_2\rangle_\perp &=\frac{\left(\mathrm{i}\left|\mathbf{k}\right|\right)^n}{4\sqrt{2}a_0^4}\bm{\xi}_{m_{2}}\cdot\bm{\epsilon}_{\left(\varkappa\right)}
\left(\mathbf{k}\right)\int_{-1}^1\mathrm{d}\eta\left(1-\eta^2\right)\eta^n \int_0^{+\infty}\mathrm{d}x\,x^{4+n}\exp\left(-\frac{3}{2}\frac{x}{a_0}\right)\\
&=\frac{1}{2\sqrt{2}}\frac{1}{a_0^4}\left(\mathrm{i}\left|\mathbf{k}\right|\right)^n\frac{1+\left(-1\right)^n}{\left(1+n\right)\left(3+n\right)}\left(\frac{3}{2a_0}\right)^{-\left(n+5\right)}\left(n+4\right)! \ \bm{\xi}_{m_{2}}\cdot\bm{\epsilon}_{\left(\varkappa\right)}\left(\mathbf{k}\right).
\end{split}
\end{equation}
Then we compute the sum over all values of $n$, according to the expression found in (\ref{eq:SeriesHere}):
\begin{equation} \label{eq:Series}
\begin{split}
\left\langle1\mathrm{s}\left|\sum_{n=0}^{\infty}\frac{\left(\mathrm{i}\mathbf{k}\cdot\hat{\mathbf{r}}\right)^n}{\left(n+1\right)!} \ \hat{\mathbf{r}}\right|2\mathrm{p}\,m_2\right\rangle_\perp &=\frac{\bm{\xi}_{m_{2}}\cdot\bm{\epsilon}_{\left(\varkappa\right)}\left(\mathbf{k}\right)}{2\sqrt{2}a_0^4}\left(\frac{3}{2a_0}\right)^{-5}
\sum_{n=0}^{\infty}\left(\mathrm{i}\left|\mathbf{k}\right|\right)^n\frac{1+\left(-1\right)^n}{\left(1+n\right)\left(3+n\right)} \left(\frac{3}{2a_0}\right)^{-n}\frac{\left(n+4\right)!}{\left(n+1\right)!}\\
&=2\sqrt{2}\left(\frac{2}{3}\right)^5a_0 \ \bm{\xi}_{m_{2}}\cdot\bm{\epsilon}_{\left(\varkappa\right)}\left(\mathbf{k}\right) \sum_{n=0}^{\infty}\left(\frac{2}{3}\left|\mathbf{k}\right|a_0\right)^{2n}\left(-1\right)^n\frac{\left(n+1\right)\left(n+2\right)}{\left(2n+1\right)}.
\end{split}
\end{equation}
It then comes in handy to notice that
\begin{equation} \label{eq:RightNotice}
\frac{\left(n+1\right)\left(n+2\right)}{\left(2n+1\right)}=\frac{1}{2}\left[\left(n+1\right)+\frac{3}{2}\left(1+\frac{1}{2n+1}\right)\right]
\end{equation}
to rewrite (\ref{eq:Series}) as
\begin{multline} \label{eq:SeriesID}
\left\langle 1\mathrm{s} \left|\sum_{n=0}^{\infty}\frac{\left(\mathrm{i}\mathbf{k}\cdot\hat{\mathbf{r}}\right)^n}{\left(n+1\right)!} \ \hat{\mathbf{r}}\right|2\mathrm{p}\,m_2\right\rangle_\perp=\sqrt{2}\left(\frac{2}{3}\right)^5a_0 \ \bm{\xi}_{m_{2}}\cdot\bm{\epsilon}_{\left(\varkappa\right)}\left(\mathbf{k}\right)\\\hspace{-10pt}\left[\frac{1}{\left[1+\left(\frac{2}{3}a_0\left|\mathbf{k}\right|\right)^2\right]^2}+\frac{3}{2}\left(\frac{1}{1+\left(\frac{2}{3}a_0\left|\mathbf{k}\right|\right)^2}+\frac{\arctan\left(\frac{2}{3}a_0\left|\mathbf{k}\right|\right)}{\frac{2}{3}a_0\left|\mathbf{k}\right|}\right)\right].
\end{multline}
\end{widetext}
Using (\ref{eq:SeriesHere}) and the following relations \cite{BetheSalpeter}
\begin{subequations} \label{eq:BohrAtom}
\begin{align}
\mathbf{d}_{\mathrm{ge}}&=e\,a_0\frac{2^{\frac{15}{2}}}{3^5}\bm{\xi}_{m_{2}},\\
\omega_{\mathrm{eg}}&=\frac{3}{8}\frac{\hbar}{m_e}\frac{1}{a_0^2}
\end{align}
\end{subequations}
we then retrieve (\ref{eq:SeriesCoupling}).

\section{$\hat{\mathbf{A}}\cdot\hat{\mathbf{p}}$ coupling \label{sec:AdotP}} 

\subsection{Dipole approximation \label{subsec:AdotPDipole}}
\begin{figure}[t]
  \begin{center}
    \includegraphics[width=1\columnwidth]{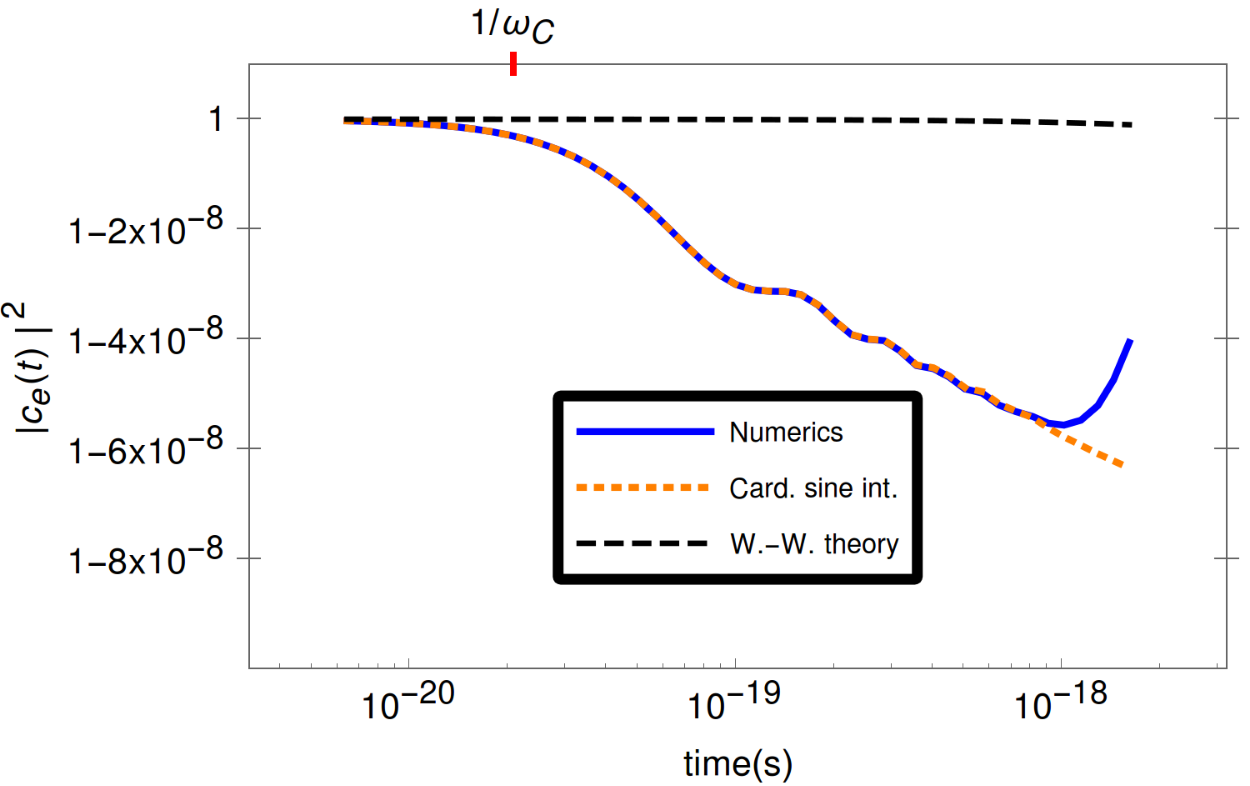}
  \end{center}
  \caption{Decay of the survival probability $\left|c_{\mathrm{e}}\left(t\right)\right|^2$ with the dipole-approximated $\hat{\mathbf{A}}\cdot\hat{\mathbf{p}}$ coupling from numerical calculations (solid blue), Wigner-Weisskopf theory (dashed black) and the truncated cardinal sine integral (\ref{eq:DipoleDecayFacchi}) (dotted orange) with $\left|c_{\mathrm{e}}\left(t\right)\right|^2=1-P_{\mathrm{decay}}\left(t\right)$. A logarithmic scale for time is used. Quasiresonant modes are discretized with $N_{\mathrm{comb}}=100$ and off-resonant modes are discretized with $N_{\mathrm{offshell}}=100$. Remember $1/\omega_{\mathrm{C}}=\SI{2.09d-20}{s}$, $1/\omega_{\mathrm{eg}}=\SI{6.45d-17}{s}$ and $1/\Gamma=\SI{1.60d-9}{s}$. \label{fig:APDS}}
\end{figure}

\begin{figure}[t]
  \begin{center}
    \includegraphics[width=1\columnwidth]{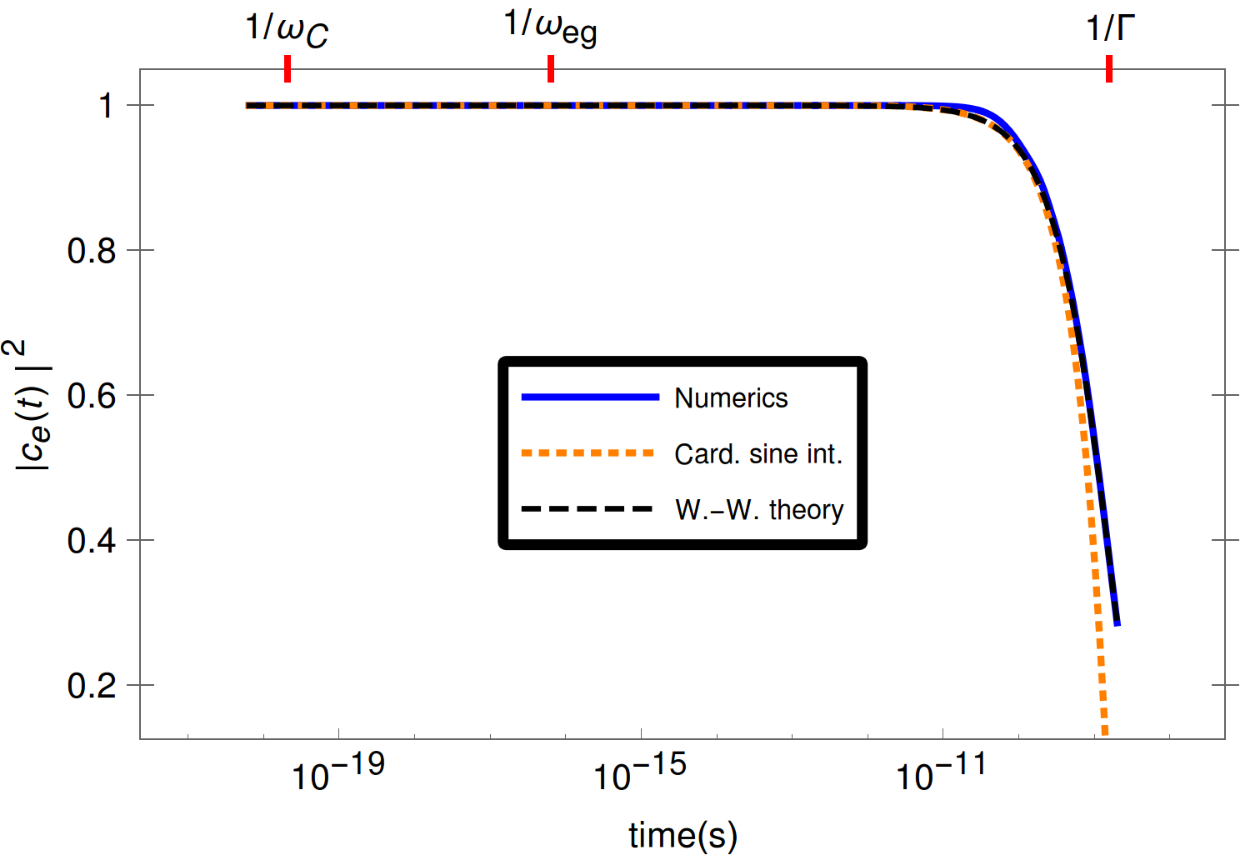}
  \end{center}
  \caption{Decay of the survival probability $\left|c_{\mathrm{e}}\left(t\right)\right|^2$ with the dipole-approximated $\hat{\mathbf{A}}\cdot\hat{\mathbf{p}}$ coupling from numerical calculations (solid blue), Wigner-Weisskopf theory (dashed black) and the truncated cardinal sine integral (\ref{eq:DipoleDecayFacchi}) (dotted orange) with $\left|c_{\mathrm{e}}\left(t\right)\right|^2=1-P_{\mathrm{decay}}\left(t\right)$. At long times numerical calculations are indistinguishable from Wigner-Weisskopf decay. A logarithmic scale for time is used. Quasiresonant modes are discretized with $N_{\mathrm{comb}}=100$ and off-resonant modes are discretized with $N_{\mathrm{offshell}}=100$. Remember $1/\omega_{\mathrm{C}}=\SI{2.09d-20}{s}$, $1/\omega_{\mathrm{eg}}=\SI{6.45d-17}{s}$ and $1/\Gamma=\SI{1.60d-9}{s}$. \label{fig:APD}}
\end{figure}

In the dipole approximation, the matrix element of the interaction Hamiltonian $\hat{\mathbf{A}}\cdot\hat{\mathbf{p}}$ is \cite{FacchiPhD}
\begin{equation} \label{eq:DipoleFacchiCoupling}
  G_\lambda\left(\mathbf{r}_0\right)=\frac{\omega_{\mathrm{eg}}}{\sqrt{2 \hbar \epsilon_0\omega_\lambda}}\  \mathbf{d}_{\mathrm{ge}}\cdot \mathbf{v}_\lambda^{T*}\left(\mathbf{r}_0\right).
\end{equation}
This time, first-order perturbation theory yields the following cardinal sine integral
\begin{multline} \label{eq:DipoleDecayFacchi}
  P_{\mathrm{decay}}\left(t\right)=\frac{t^2 \left|\mathbf{d}_{\mathrm{ge}}\right|^2}{6\pi^2\epsilon_0\hbar c^3} \ \omega_{\mathrm{eg}}^2 \\
  \int_0^{c\,K_{\mathrm{C}}}\mathrm{d}\omega \ \omega \ \sin\!\mathrm{c}^2\left(\left(\omega_{\mathrm{eg}}-\omega\right)\frac{t}{2}\right).
\end{multline}
The same perturbation method as that exposed in Sect.~\ref{sec:EdotXDipole} is used. The frequencies $\omega_\lambda$ and couplings $G_\lambda\left(\mathbf{r}_0\right)$ are defined in App.~\ref{subsubsec:AdotPDipoleAppdx}. Results with this coupling are shown in Figs.~\ref{fig:APDS} (short times) and \ref{fig:APD} (long times). A comparison of Figs.~\ref{fig:APDS} and \ref{fig:Final} shows that, even in the dipole approximation, the Zeno effect obtained is very much reduced in the $\hat{\mathbf{A}}\cdot\hat{\mathbf{p}}$ coupling compared to what it is in the $\hat{\mathbf{E}}\cdot\hat{\mathbf{x}}$ coupling.


\subsection{Exact $\hat{\mathbf{A}}\cdot\hat{\mathbf{p}}$ coupling \label{subsec:AdotPExact}}
\begin{figure}[t]
  \begin{center}
    \includegraphics[width=1\columnwidth]{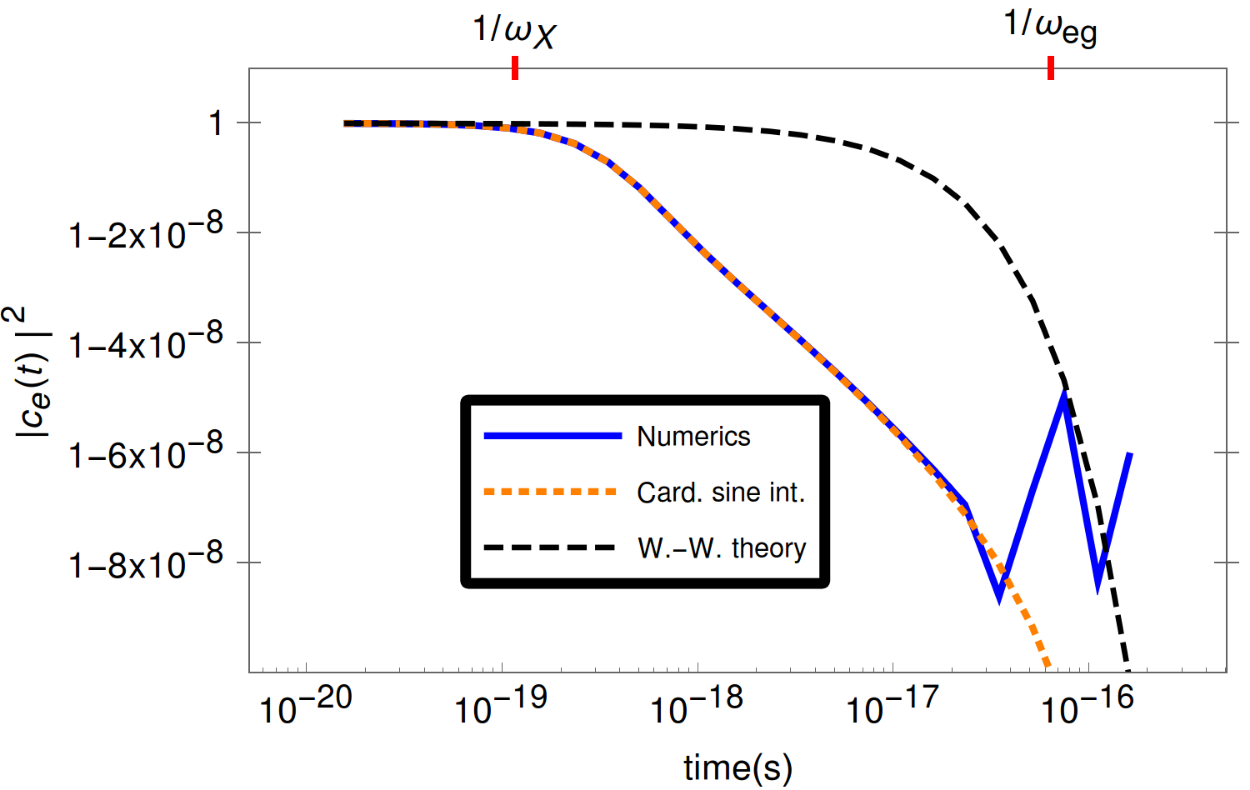}
  \end{center}
  \caption{Decay of the survival probability $\left|c_{\mathrm{e}}\left(t\right)\right|^2$ with the exact $\hat{\mathbf{A}}\cdot\hat{\mathbf{p}}$ coupling from numerical calculations (solid blue), Wigner-Weisskopf theory (dashed black) and the cardinal sine integral (\ref{eq:DecayFacchi}) (dotted orange) with $\left|c_{\mathrm{e}}\left(t\right)\right|^2=1-P_{\mathrm{decay}}\left(t\right)$. Numerical calculations are indistinguishable from (\ref{eq:DecayFacchi}). A logarithmic scale for time is used. Quasiresonant modes are discretized with $N_{\mathrm{comb}}=100$ and off-resonant modes are discretized with $N_{\mathrm{offshell}}=100$. Remember $1/\omega_{\mathrm{X}}=\SI{1.18d-19}{s}$, $1/\omega_{\mathrm{eg}}=\SI{6.45d-17}{s}$ and $1/\Gamma=\SI{1.60d-9}{s}$. \label{fig:TrueZeno}}
\end{figure}

\begin{figure}[t]
  \begin{center}
    \includegraphics[width=1\columnwidth]{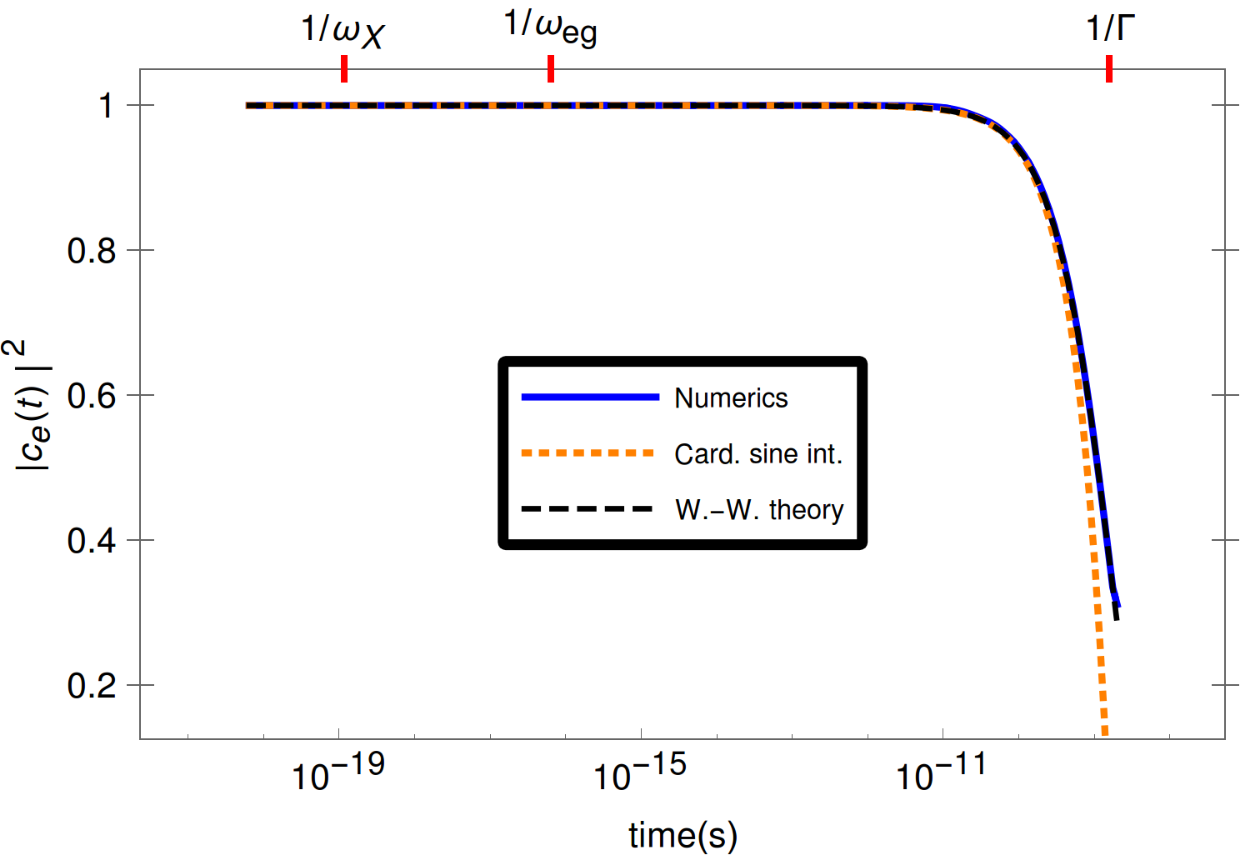}
  \end{center}
  \caption{Decay of the survival probability $\left|c_{\mathrm{e}}\left(t\right)\right|^2$ with the exact $\hat{\mathbf{A}}\cdot\hat{\mathbf{p}}$ coupling from numerical calculations (solid blue), Wigner-Weisskopf theory (dashed black) and the cardinal sine integral (\ref{eq:DecayFacchi}) (dotted orange) with $\left|c_{\mathrm{e}}\left(t\right)\right|^2=1-P_{\mathrm{decay}}\left(t\right)$. At long times numerical calculations are indistinguishable from Wigner-Weisskopf decay. A logarithmic scale for time is used. Quasiresonant modes are discretized with $N_{\mathrm{comb}}=100$ and off-resonant modes are discretized with $N_{\mathrm{offshell}}=100$. Remember $1/\omega_{\mathrm{X}}=\SI{1.18d-19}{s}$, $1/\omega_{\mathrm{eg}}=\SI{6.45d-17}{s}$ and $1/\Gamma=\SI{1.60d-9}{s}$. \label{fig:EndlichGamma}}
\end{figure}

When going beyond the dipole approximation, one can derive a more exact expression for the matrix element, namely \cite{FacchiPhD}
\begin{equation} \label{eq:FacchiCoupling}
  G_\lambda\left(\mathbf{r}_0\right)=\frac{\omega_{\mathrm{eg}}}{\sqrt{2 \hbar \epsilon_0\omega_\lambda}}\  \mathbf{d}_{\mathrm{ge}}\cdot \mathbf{v}_\lambda^{T*}\left(\mathbf{r}_0\right)\frac{1}{\left[1+\left(\frac{\omega_\lambda}{\omega_{\mathrm{X}}}\right)^2\right]^2}
\end{equation}
with natural cutoff frequency $\omega_{\mathrm{X}}=\left(3c\right)/\left(2a_0\right)$. A mode discretization similar to that used in Sect.~\ref{sec:EdotXDipole}, and the important results given in App.~\ref{subsubsec:AdotPExactAppdx}. This time, first-order perturbation theory yields the following cardinal sine integral
\begin{multline} \label{eq:DecayFacchi}
  P_{\mathrm{decay}}\left(t\right)=\frac{t^2 \left|\mathbf{d}_{\mathrm{ge}}\right|^2 }{6\pi^2\epsilon_0\hbar c^3}\omega_{\mathrm{eg}}^2\\
  \int_0^{\infty}\mathrm{d}\omega \ \frac{\omega}{\left[1+\left(\frac{\omega}{\omega_{\mathrm{X}}}\right)^2\right]^4} \ \sin\!\mathrm{c}^2\left(\left(\omega_{\mathrm{eg}}-\omega\right)\frac{t}{2}\right).
\end{multline}
The same perturbation method as that exposed in Sect.~\ref{sec:EdotXDipole} is used, with the following modifications. The frequencies $\omega_\lambda$ and couplings $G_\lambda\left(\mathbf{r}_0\right)$ are defined in App.~\ref{subsubsec:AdotPExactAppdx}. Results with this coupling are shown in Figs.~\ref{fig:TrueZeno} (short times) and \ref{fig:EndlichGamma} (long times). Figs.~\ref{fig:EX} through \ref{fig:EndlichGamma} show that the strong decay of the survival probability observed on Fig.~\ref{fig:Final} for very short times is an incorrect prediction. This highlights the fact that, if the dipole approximation is to be used at all, this should be done with much care. Note that for short times our perturbative approach fits the analytical results derived in \cite{FacchiPascazio}, by a ``tour de force'' which allowed the authors of \cite{FacchiPascazio} to estimate departures from exponential decay for short but also very long times in an exact fashion.

It is worth noting that while Facchi and Pascazio considered \cite{FacchiPascazio} the ratio between the Zeno time $\tau_{\mathrm{Z}}$ and the lifetime $\tau_{\mathrm{E}}$ of the excited level, where $P_{\mathrm{surv}}\left(t\right)\underset{t\rightarrow0}{\sim}1-\left(t/\tau_{\mathrm{Z}}\right)^2$ and $t_{\mathrm{E}}\equiv1/\Gamma$, as the relevant parameter concerning the experimental observability of the Zeno region, we argue that the relevant ratio is that between the ``cutoff time'' $\tau_{\mathrm{X}}$ and the Zeno time $\tau_{\mathrm{Z}}$. The cutoff time is simply defined as $\tau_{\mathrm{X}}=1/\omega_{\mathrm{X}}$ with $\omega_{\mathrm{X}}$ the cutoff frequency of the atom-field interaction. Indeed, the response of the electromagnetic modes is no longer coherent after a time of order $\tau_{\mathrm{X}}$ (as can be seen for instance on Fig.~\ref{fig:WWVSSC} where the cutoff frequency is $\omega_{\mathrm{C}}$). This means that after $\tau_{\mathrm{X}}$, one exits the Zeno regime in which the survival probability decays quadratically. With this in mind, we take an interest in the scaling properties of the ratio $\tau_{\mathrm{X}}/\tau_{\mathrm{Z}}$. For hydrogen-like atoms with $Z$ protons, the Zeno time scales, as noted in \cite{FacchiPascazio}, like $Z^{-2}$. For such systems, since the Bohr radius scales like $Z^{-1}$, the cutoff frequency scales like $Z$, and the cutoff time like $Z^{-1}$, so that the ratio $\tau_{\mathrm{X}}/\tau_{\mathrm{Z}}$ scales like $Z$, while Facchi and Pascasio's parameter ($\tau_{\mathrm{Z}}/\tau_{\mathrm{E}}$) scales like $Z^2$.


\section{Conclusion and open questions \label{sec:Ccl}}

We scrutinized the onset of the exponential decay of a quantum dipole in the Zeno regime. We focused on the role played by  excitations of the off-shell electro-magnetic modes that are usually neglected. Our main result is that, before an exponential behaviour is manifest and even before the linear in time decrease of the survival probability characteristic of the Fermi golden rule appears, a sudden decrease occurs for very short times, quadratic in time,  which is directly related to the so-called quantum Zeno effect.

\begin{figure}[t]
  \begin{center}
    \includegraphics[width=1\columnwidth]{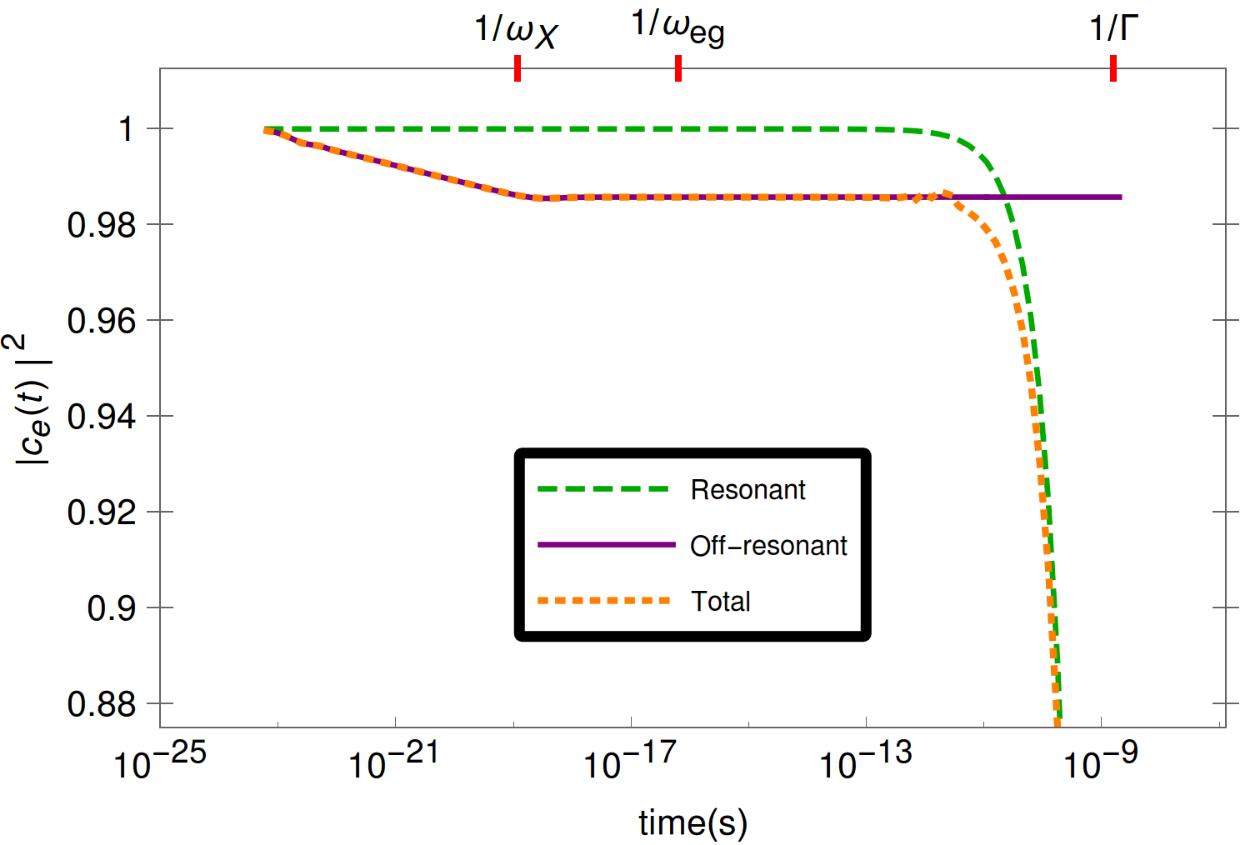}
  \end{center}
  \caption{Decay of the survival probability $\left|c_{\mathrm{e}}\left(t\right)\right|^2$ for the exact $\hat{\mathbf{E}}\cdot\hat{\mathbf{x}}$ coupling if one takes into account either only resonant electromagnetic modes (dashed green), only off-resonant modes (solid violet) and all modes (dotted orange). Remember $1/\omega_{\mathrm{X}}=\SI{1.18d-19}{s}$, $1/\omega_{\mathrm{eg}}=\SI{6.45d-17}{s}$ and $1/\Gamma=\SI{1.60d-9}{s}$. \label{fig:Ultra}}
\end{figure}

As can be seen in Fig.~\ref{fig:Ultra} (where the onshell and offshell contributions to the decay of the survival probability as expressed by the integral (\ref{eq:SeriesDecay}) are plotted separately), the offshell modes play a crucial role in the Zeno regime. They are activated very quickly but after a time of the order of the inverse of the cut-off frequency they do no longer respond coherently and their contribution to the decay vanishes. The linear behavior predicted by the Fermi golden rule occurs much later and is due to onshell contributions only.

As a by-product of our analysis, we developed a perturbative scheme that interpolates between the Zeno (parabolic) regime and the W-W (exponential) regime, passing through the Fermi (linear) regime. In the case of the $2\text{p}-1\text{s}$ transition we cover a very large interval of time (from 10$^{-19}$ to 10$^{-9}$ second). Our results fit with the exact solution of Facchi and Pascasio \cite{FacchiPascazio} if we treat the $\hat{\mathbf{A}}\cdot\hat{\mathbf{p}}$ coupling without approximation, but our perturbative approach also makes it possible to treat the exact $\hat{\mathbf{E}}\cdot\hat{\mathbf{x}}$ coupling as well as the $\hat{\mathbf{A}}\cdot\hat{\mathbf{p}}$ and $\hat{\mathbf{E}}\cdot\hat{\mathbf{x}}$ coupling in the dipolar approximation. 

It is natural to ask whether the Zeno effect predicted in our paper is likely to be observed experimentally. The effect that we predict, in the case of the $1\mathrm{s}$-$2\mathrm{p}$ transition is, indeed, very small and does not occur during a measurable time period (typically, of the order of an attosecond). This explains why our counter-intuitive predictions have not yet been verified experimentally. This verification,
however, is not impossible in principle, but requires
\begin{enumerate}
\item{to measure the change of population of the emitters carefully.}
\item{to investigate other unstable quantum systems (magnetic dipoles \cite{Norge,BriceMagnetic}, solid state or NMR qubits, Rydberg states \cite{HarocheRaimond} and so on). The reason is that the duration and intensity of the Zeno effect is strongly dependent on the off-shell coupling parameters while the W-W and Fermi golden rule regimes are dominated by the on-shell coupling. Up to now, we limited ourselves to the study of a single transition (the $1\mathrm{s}-2\mathrm{p}$ transition for atomic hydrogen) but we are convinced that there exist in Nature other transitions (quantum jumps) where the particular type of Zeno effect predicted by us (collective response of the off-shell modes at zero temperature) is enhanced. Magnetic dipolar transitions \cite{Norge,BriceMagnetic} for instance are characterized by very long life times (in some cases a million times longer than the $1\mathrm{s}-2\mathrm{p}$ transition), which suggests that the Zeno time could approach the picosecond, a temporal regime which can be probed by sophisticated pump-probe techniques such as the so-called ultrafast transient absorption spectroscopy in which very quick responses (of the order of 200 fs) are reachable experimentally. }
\end{enumerate}

Another interesting question concerns causality. It is known that, in accordance with Hegerfeldt's theorem, even if we stick to the W-W approximation, a non-zero component of the single photon wave function will be present outside from the light cone expanding centered in spacetime on the atom at time $t=0$ \cite{FermiCausal,Shirokov,HegerfeldtFermi}. The reason therefor is that only positive spatial frequencies are present in the spectrum of this wave function, so that, in virtue of Paley-Wiener theorem, the domain of the wave function cannot be bounded. We investigated the departures from causality in the W-W regime in another paper and found that they are extremely small \cite{DDDNVZHaroche}.

One could wonder whether more significant violations of Einsteinian causality might be present in the Zeno regime. In order to tackle this question, it is necessary to have at one's disposal extremely accurate estimates of the wave function of the emitted photon. This is possible in our approach which provides an accurate estimate of the solution $c_{\mathrm{e}}\left(t\right)$ of the equations of motion (\ref{eq:ExactEQ}) that interpolates between the Zeno time and the lifetime of the dipole. This suffices in order to compute the field emitted by the atom. Indeed, the formal solution of (\ref{eq:ExactEQg}) reads
\begin{equation}\label{eq:FormalSol}
  c_{\mathrm{g},\lambda}\left(t\right) =   G_\lambda\left(\mathbf{r}_0\right) \ \int_0^t c_{\mathrm{e}}(t') \ \mathrm{e}^{\mathrm{i} (\omega_\lambda-\omega_{\mathrm{eg}})  t'} \ \mathrm{d} t'.
\end{equation}
The information about the single-photon field (and in particular of its Glauber wave function) emitted by the atom is entirely encoded in the coefficients $c_{\mathrm{g},\lambda}\left(t\right)$. We expect therefore to be able to study in depth possible violations of Einsteinian causality for very short times. Of course, we performed several approximations in our approach (RWA for instance) and it remains an open question to know whether violations of causality are due to the approximations -- as suggested in \cite{PassanteVirtual} -- or whether they are unavoidable features of QED. These questions are complex but at least our approach makes it possible to tackle the problem in an original manner.


\appendix*

\section{Mode discretization: energy slice ladder operators \label{sec:DoS}} 

\subsection{Position of the problem \label{subsec:AppIntro}} 
In the present appendix we show how to discretize the continuous modes of free space through the introduction of creation and annihilation photon operators for energy slices. We are interested in the spontaneous emission of light in a vacuum and hence our starting point will feature a continuous infinity of modes. Hence we first want to adapt the equations of Sect.~\ref{subsec:QED} to this situation. In Sect.~\ref{subsec:NoTimeTime} we gave part of the prescription for switching from a discrete sum over arbitrary modes to a continuous sum over the free space modes. This was done by writing
\begin{equation} \label{eq:DiscretetoContinuousApp}
  \frac{1}{V}\sum_\lambda\rightarrow\sum_{\varkappa=1}^2\int\frac{\mathrm{d}^3k}{\left(2\pi\right)^3}
\end{equation}
where $\varkappa$ are the two possible photon polarizations, $V$ is the initial quantization volume and the modal functions are given, in the Coulomb gauge, by
\begin{equation} \label{eq:VacuumModesApp}
  \mathbf{v}_\lambda^T\left(\mathbf{r}\right)\rightarrow\mathbf{v}_\varkappa^T\left(\mathbf{k},\mathbf{r}\right)=\frac{\mathrm{e}^{\mathrm{i}\mathbf{k}\cdot\mathbf{r}}}{\sqrt{V}} \ \bm{\epsilon}_{\left(\varkappa\right)}\left(\mathbf{k}\right).
\end{equation}
It will come in handy here to choose linear polarization vectors as a basis. These are given by

\begin{equation} \label{eq:UnitSphere}
  \arraycolsep=2.5pt
  \begin{array}{lclrrclrclr}
    \bm{\epsilon}_{\left(1\right)}\left(\mathbf{k}\right)&=&\cos\theta\cos\varphi&\mathbf{e}_x&+&\cos\theta\sin\varphi&\mathbf{e}_y&-&\sin\theta&\mathbf{e}_z,\\
    \bm{\epsilon}_{\left(2\right)}\left(\mathbf{k}\right)&=&-\sin\varphi&\mathbf{e}_x&+&\cos\varphi&\mathbf{e}_y&&&
  \end{array}
\end{equation}
where $\theta$ and $\phi$ are the spherical coordinates of the wave vector $\mathbf{k}$ with respect to an arbitrary cartesian frame (here, we will take $\mathbf{e}_z$ in the direction of $\mathbf{d}_{\mathrm{ge}}$. A presciption for going over to the continuous case for ladder operators needs to be added. It consists in the substitution
\begin{equation} \label{eq:AppOnly}
  \left[\hat{a}_\lambda,\hat{a}_{\lambda'}^\dagger\right]=\delta_{\lambda\lambda'}\rightarrow\left[\hat{a}_\varkappa\left(\mathbf{k}\right),\hat{a}_\zeta^\dagger\left(\mathbf{q}\right)\right]=\left(2\pi\right)^3\delta\left(\mathbf{k}-\mathbf{q}\right)\delta_{\varkappa\zeta}.
\end{equation}
Now we want to be able to perform numerical computations, and hence have to discretize this continuum in some way. 

\subsection{Results \label{subsec:SliceThemUp}} 

\subsubsection{Detailed treatment in the case of the dipolar $\hat{\mathbf{E}}\cdot\hat{\mathbf{x}}$ coupling \label{subsubsec:EdotXDipoleAppdx}} 

We then rewrite the rotating interaction Hamiltonian (\ref{eq:Resonant}) with the use of (\ref{eq:DiscretetoContinuousApp}), (\ref{eq:VacuumModesApp}) and, in the second step, (\ref{eq:UnitSphere}) as
\begin{widetext}
\begin{equation} \label{eq:IntInt}
  \begin{aligned} [b]
    \hat{H}_I^R&=-\frac{\mathrm{i}}{\left(2\pi\right)^3}
    \sum_{\varkappa=1}^2\int_0^{2\pi}\mathrm{d}\varphi\int_0^\pi
    \mathrm{d}\theta \ \sin\theta \ \sqrt{\frac{\hbar c}{2\epsilon_0}}
    \int_0^{\infty}\mathrm{d}k \ k^{\frac{5}{2}} \ \left[\hat{a}_\varkappa\left(k,\theta,\varphi\right)\bm{\epsilon}_{\left(\varkappa\right)}\left(\theta,\varphi\right)\cdot\mathbf{d}_{\mathrm{ge}}^*|\mathrm{e}\rangle\langle\mathrm{g}|-\text{H.c.}\right]\\
    &=-\frac{\mathrm{i}}{\left(2\pi\right)^3} \ \sqrt{\frac{\hbar c}{2\epsilon_0}} \ \left|\mathbf{d}_{\mathrm{ge}}\right|\int_0^{2\pi}\mathrm{d}\varphi\int_0^\pi\mathrm{d}\theta \ \sin\theta \ \int_0^{\infty}\mathrm{d}k \ k^{\frac{5}{2}} \ \left(-\sin\theta\right)\left[\hat{a}_1\left(k,\theta,\varphi\right)\mid\!\mathrm{e}\rangle\langle\mathrm{g}\!\mid-\hat{a}_1^\dagger\left(k,\theta,\varphi\right)\mid\!\mathrm{g}\rangle\langle\mathrm{e}\!\mid\right]
  \end{aligned}
\end{equation}
where we took $\mathbf{d}_{\mathrm{ge}}$ as the reference axis for the polar coordinate $\theta$, and used the fact that, for the $2 p- 1s$ transition in atomic hydrogen, the dipole matrix element is real \cite{BetheSalpeter}. With the prescription to obtain an effective coupling given below (\ref{eq:Decoupled}) in mind, we introduce the spherically averaged creation operators
\begin{equation} \label{eq:SphericalA}
    \int_0^{2\pi}\mathrm{d}\varphi \int_0^\pi\mathrm{d}\theta \ \sin^2\theta \ \hat{a}_1\left(k,\theta,\varphi\right) \equiv\hat{a}_{\mathrm{sph}}\left(k\right)\left(\int_0^{2\pi}\mathrm{d}\varphi\int_0^\pi
    \mathrm{d}\theta \ \sin^3\theta\right)^{\frac{1}{2}}=2\sqrt{\frac{2\pi}{3}}\hat{a}_{\mathrm{sph}}\left(k\right).
\end{equation}
\end{widetext}
In the leftmost member of (\ref{eq:SphericalA}), $\sin^2\theta$ is the product of a first sine coming from the differential solid angle element, with another one coming from the coupling (see (\ref{eq:IntInt}). According to our prescription, the coupling of one particular mode (here, with angular orientation $\left(\theta,\varphi\right)$) should be squared to obtain, after summing over a given ensemble of modes and taking the square root, the effective coupling constant of that ensemble, hence the third power of $\sin\theta$ in the central term of (\ref{eq:SphericalA}). With these new operators, the interaction Hamiltonian (\ref{eq:IntInt}) can be rewritten
\begin{multline} \label{eq:IntSphInt}
  \hat{H}_I^R=\frac{\mathrm{2i}}{\left(2\pi\right)^3} \ \sqrt{\frac{\hbar c\pi}{3\epsilon_0}} \ \left|\mathbf{d}_{\mathrm{ge}}\right| \ \int_0^{\infty}\mathrm{d}k \ k^{\frac{5}{2}}\\
  \left[\hat{a}_{\mathrm{sph}}\left(k\right) \ \mid\!\mathrm{e}\rangle\langle\mathrm{g}\!\mid-\hat{a}_{\mathrm{sph}}^\dagger\left(k\right) \ \mid\!\mathrm{g}\rangle\langle\mathrm{e}\!\mid\right].
\end{multline}
It can be checked from (\ref{eq:AppOnly}) and (\ref{eq:SphericalA}) that the spherically integrated ladder operators obey the following commutation relation:
\begin{equation} \label{eq:SphComm}
  \left[\hat{a}_{\mathrm{sph}}\left(k\right),\hat{a}_{\mathrm{sph}}^\dagger\left(q\right)\right]=\frac{\left(2\pi\right)^3}{kq}\delta\left(k-q\right).
\end{equation}
We further define
\begin{equation} \label{eq:AverageA}
  \begin{aligned} [b]
    \mathrm{i}\left(2\pi\right)^{-\frac{3}{2}}\int_{k_\lambda}^{k_{\lambda+1}}\mathrm{d}k\,k^{\frac{5}{2}}\hat{a}_{\mathrm{sph}}\left(k\right)&\equiv\hat{a}_{\mathrm{avg}}^\lambda\left(\int_{k_\lambda}^{k_{\lambda+1}}\mathrm{d}k\,k^3\right)^{\frac{1}{2}}\\
    &=\left(\frac{k_{\lambda+1}^4-k_\lambda^4}{4}\right)^{\frac{1}{2}}\hat{a}_{\mathrm{avg}}^\lambda
  \end{aligned}
\end{equation}
to rewrite (\ref{eq:IntSphInt}) as
\begin{multline} \label{eq:IntAvgSum}
  \hat{H}_I^R=\frac{\mathrm{2}}{\left(2\pi\right)^{\frac{3}{2}}}\sqrt{\frac{\hbar c\pi}{3\epsilon_0}}\left|\mathbf{d}_{\mathrm{ge}}\right|\sum_\lambda\left(\frac{k_{\lambda+1}^4-k_\lambda^4}{4}\right)^{\frac{1}{2}}\\
  \left[\hat{a}_{\mathrm{avg}}^\lambda\mid\!\mathrm{e}\rangle\langle\mathrm{g}\!\mid+\left(\hat{a}_{\mathrm{avg}}^\lambda\right)^\dagger\mid\!\mathrm{g}\rangle\langle\mathrm{e}\!\mid\right]
\end{multline}
where $\hat{a}_{\mathrm{avg}}^\lambda\equiv\hat{a}_{\mathrm{avg}}^{k_\lambda k_{\lambda+1}}$ defined through (\ref{eq:AverageA}) and the intervals $\left[k_\lambda,k_{\lambda+1}\right]$ are taken to have empty intersections and $\left[0,+\infty\right[$ as their union. It can be checked from (\ref{eq:SphComm}) and (\ref{eq:AverageA}) that the energy slice ladder operators obey the following commutation relation:
    \begin{equation} \label{eq:AvgComm}
      \left[\hat{a}_{\mathrm{avg}}^\lambda,\hat{a}_{\mathrm{avg}}^{\zeta\dagger}\right]=\delta_{\lambda\zeta}.
    \end{equation}
From this we can compute the matrix elements of the interaction Hamiltonian between the excited state $\mid\!\mathrm{e},0\rangle$ and the energy slice field state $\left(\hat{a}_{\mathrm{avg}}^\lambda\right)^\dagger\mid\!\mathrm{g},0\rangle$, which reads
    \begin{equation} \label{eq:IntMatrixElement}
      \langle\mathrm{e},0|\hat{H}_I^R \left(\hat{a}_{\mathrm{avg}}^\lambda\right)^\dagger|\mathrm{g},0\rangle=\frac{\mathrm{1}}{2\pi}\sqrt{\frac{\hbar c}{6\epsilon_0}}\left|\mathbf{d}_{\mathrm{ge}}\right|\left(k_{\lambda+1}^4-k_\lambda^4\right)^{\frac{1}{2}}.
    \end{equation}
    For a complete description we also need to compute the matrix elements of the free field Hamiltonian for these energy slice field states. The electromagnetic field Hamiltonian (\ref{eq:FieldHamilton}) is rewritten, in free space,
    \begin{multline} \label{eq:FreeFreeField}
      \hat{H}_R=\frac{\hbar c}{\left(2\pi\right)^3}\sum_{\varkappa=1}^2\int_0^{2\pi} \mathrm{d}\varphi \ \int_0^\pi\mathrm{d}\theta \ \sin\theta \\
      \int_0^{\infty} \mathrm{d}k \ k^3 \ \hat{a}_\varkappa^\dagger\left(k,\theta,\varphi\right)\hat{a}_\varkappa\left(k,\theta,\varphi\right).
    \end{multline}
    As seen from (\ref{eq:UnitSphere}), electromagnetic modes polarized orthogonally to the atomic dipole vector $\mathbf{d}_{\mathrm{ge}}$ are not coupled to the atom and thus evolve freely. It is thus only useful to consider
    \begin{equation} \label{eq:WhatWeWant}
      \langle\mathrm{g},0\!\mid\hat{a}_{\mathrm{avg}}^\lambda\,\hat{H}_R\,\hat{a}_{\mathrm{avg}}^{\zeta\dagger}\mid\!\mathrm{g},0\rangle.
    \end{equation}
    To that end it is best to first compute the commutator
    \begin{multline} \label{eq:Buck}
      \left[\hat{a}_1\left(q,\theta,\varphi\right),\left(\hat{a}_{\mathrm{avg}}^\lambda\right)^\dagger\right]=-\left(2\pi\right)^{\frac{3}{2}}\frac{\mathrm{i}}{\left(k_{\lambda+1}^4-k_\lambda^4\right)^{\frac{1}{2}}}\left(\frac{3}{2\pi}\right)^{\frac{1}{2}}\\
      \sin\theta \ \sqrt{q} \ \left[\Theta\left(q-k_\lambda\right)-\Theta\left(q-k_{\lambda+1}\right)\right]
    \end{multline}
    where $k_\lambda$ and $k_{\lambda+1}$ are the bounds of the $i^{\text{th}}$ interval (energy slice). This can be seen to be true from (\ref{eq:AppOnly}), (\ref{eq:SphericalA}) and (\ref{eq:AverageA}). Noticing that since $k_{\lambda+1}>k_\lambda$, we have
\begin{widetext}
    \begin{equation} \label{eq:ThetaGalore}
      \begin{aligned} [b]
        \left[\Theta\left(q-k_\lambda\right)-\Theta\left(q-k_{\lambda+1}\right)\right]^2&=\Theta\left(q-k_\lambda\right)+\Theta\left(q-k_{\lambda+1}\right)-2\Theta\left(q-k_\lambda\right)\Theta\left(q-k_{\lambda+1}\right)\\
        &=\Theta\left(q-k_\lambda\right)+\Theta\left(q-k_{\lambda+1}\right)-2\Theta\left(q-k_{\lambda+1}\right)=\Theta\left(q-k_\lambda\right)-\Theta\left(q-k_{\lambda+1}\right),
      \end{aligned}
    \end{equation}
 we compute (\ref{eq:WhatWeWant}) as follows:
    \begin{equation} \label{eq:FieldMatrixElement}
      \begin{aligned} [b]
        \left\langle\mathrm{g},0\left|\hat{a}_{\mathrm{avg}}^\lambda\,\hat{H}_R\,\hat{a}_{\mathrm{avg}}^{\zeta\dagger}\right|\mathrm{g},0\right\rangle&=\frac{3 \hbar c}{2\pi} \ \delta_{\lambda\zeta} \ \int_0^{2\pi}\mathrm{d}\varphi\int_0^\pi\mathrm{d}\theta\sin^3\theta \int_0^{\infty}\mathrm{d}k \ \frac{k^4}{k_{\lambda+1}^4-k_\lambda^4}\left(\Theta\left(q-k_\lambda\right)-\Theta\left(q-k_{\lambda+1}\right)\right)\\ 
        &=\frac{3\hbar c}{2\pi} \ \delta_{\lambda\zeta} \  \int_0^{2\pi}\mathrm{d}\varphi\int_0^\pi\mathrm{d}\theta\sin^3\theta \ \int_{k_\lambda}^{k_{\lambda+1}}\mathrm{d}k\,\frac{k^4}{k_{\lambda+1}^4-k_\lambda^4}\\
        &= 4\hbar c \ \delta_{\lambda\zeta} \ \int_{k_\lambda}^{k_{\lambda+1}}\mathrm{d}k\,\frac{k^4}{k_{\lambda+1}^4-k_\lambda^4}=\frac{4}{5} \ \hbar c \ \delta_{\lambda\zeta} \ \frac{k_{\lambda+1}^5-k_\lambda^5}{k_{\lambda+1}^4-k_\lambda^4}.
      \end{aligned}
    \end{equation}
\end{widetext}
Notice that if one writes $k_{\lambda+1}-k_\lambda\equiv\Delta k_\lambda$, and lets $\Delta k_\lambda\rightarrow0$, then (\ref{eq:FieldMatrixElement}) tends towards $\hbar c\,k_\lambda\simeq\hbar c\,k_{\lambda+1}$. The matrix elements (\ref{eq:IntMatrixElement}) and (\ref{eq:FieldMatrixElement}) are the ones we use for our numerical calculations. In the body of the article we denote them as $\langle\mathrm{e},0\!\mid\hat{H}_I^R\,\left(\hat{a}_{\mathrm{avg}}^\lambda\right)^\dagger\mid\!\mathrm{g},0\rangle\equiv-\mathrm{i}\hbar G_\lambda^*\left(\mathbf{r}_0\right)$ for the coupling matrix (see (\ref{eq:Resonant}) in the text) element and $\langle\mathrm{g},0\!\mid\hat{a}_{\mathrm{avg}}^\lambda\,\hat{H}_R\,\left(\hat{a}_{\mathrm{avg}}^\lambda\right)^\dagger\mid\!\mathrm{g},0\rangle\equiv\hbar\omega_\lambda$ for the diagonal field energy matrix elements.
    
\subsubsection{Results for the exact $\hat{\mathbf{E}}\cdot\hat{\mathbf{x}}$ coupling \label{subsubsec:EdotXExactAppdx}}

For the exact $\hat{\mathbf{E}}\cdot\hat{\mathbf{x}}$ coupling (\ref{eq:FacchiCoupling}) we define 
\begin{multline} \label{eq:FRelief}
F\left(k\right)=\frac{k^{\frac{1}{2}}}{4} \times\\
\left[\frac{1}{\left[1+\left(\frac{k}{k_{\mathrm{X}}}\right)^2\right]^2}+\frac{3}{2}\left(\frac{1}{1+\left(\frac{k}{k_{\mathrm{X}}}\right)^2}+\frac{\arctan\left(\frac{k}{k_{\mathrm{X}}}\right)}{\frac{k}{k_{\mathrm{X}}}}\right)\right]
\end{multline}
and obtain
\begin{multline} \label{eq:IntMatrixElementFSeries}
\langle\mathrm{e},0|\hat{H}_I^R\left(\hat{a}_{\mathrm{avg}}^\lambda\right)^\dagger|\mathrm{g},0\rangle=\frac{\mathrm{1}}{2\pi} \ \sqrt{\frac{\hbar c}{3\epsilon_0}} \ \left|\mathbf{d}_{\mathrm{ge}}\right| \ \omega_{\mathrm{eg}}\\
\left(I^{\left(4\right)}\left(k_{\lambda+1}\right)-I^{\left(4\right)}\left(k_\lambda\right)\right)^{-\frac{1}{2}}
\end{multline}
    where
    \begin{equation} \label{eq:I4}
      I^{\left(4\right)}\left(k_{\lambda+1}\right)-I^{\left(4\right)}\left(k_\lambda\right)\equiv\int_{k_\lambda}^{k_{\lambda+1}}\mathrm{d}k \ k^2 \ F^2\left(k\right)
    \end{equation}
with $\hat{a}_{\mathrm{avg}}^\lambda\equiv\hat{a}_{\mathrm{avg}}^{k_\lambda k_{\lambda+1}}$ defined through
\begin{multline} \label{eq:AverageAEX}
    \mathrm{i}\left(2\pi\right)^{-\frac{3}{2}}\  \ \int_{k_\lambda}^{k_{\lambda+1}}\mathrm{d}k \ k^2 \ F\left(k\right)\ \hat{a}_{\mathrm{sph}}\left(k\right)\\ \equiv \hat{a}_{\mathrm{avg}}^\lambda\left(\int_{k_\lambda}^{k_{\lambda+1}}\mathrm{d}k \ k^2 \ F^2\left(k\right)\right)^{\frac{1}{2}}.
\end{multline}
 We also find that
        \begin{equation} \label{eq:FieldMatrixElementFSeries}
          \langle\mathrm{g},0\!\mid\hat{a}_{\mathrm{avg}}^\lambda\,\hat{H}_R\,\hat{a}_{\mathrm{avg}}^{\zeta\dagger}\mid\!\mathrm{g},0\rangle=\hbar c\,\delta_{\lambda\zeta}\frac{I^{\left(5\right)}\left(k_{\lambda+1}\right)-I^{\left(5\right)}\left(k_\lambda\right)}{I^{\left(4\right)}\left(k_{\lambda+1}\right)-I^{\left(4\right)}\left(k_\lambda\right)}
        \end{equation}
        where
        \begin{equation} \label{eq:I5}
            I^{\left(5\right)}\left(k_{\lambda+1}\right)-I^{\left(5\right)}\left(k_\lambda\right)\equiv\int_{k_\lambda}^{k_{\lambda+1}}\mathrm{d}k \ k^3 \ F^2\left(k\right).
        \end{equation}
        As far as we know the integrals (\ref{eq:I4}) and (\ref{eq:I5}) cannot be solved by hand and we compute them numerically. The matrix elements (\ref{eq:IntMatrixElementFSeries}) and (\ref{eq:FieldMatrixElementFSeries}) are the ones we use for our numerical calculations. In the body of the article we denote them as $\langle\mathrm{e},0|\hat{H}_I^R\,\left(\hat{a}_{\mathrm{avg}}^\lambda\right)^\dagger|\mathrm{g},0\rangle\equiv-\mathrm{i}\hbar G_\lambda^*\left(\mathbf{r}_0\right)$ for the coupling matrix element and $\langle\mathrm{g},0|\hat{a}_{\mathrm{avg}}^\lambda\,\hat{H}_R\,\left(\hat{a}_{\mathrm{avg}}^\lambda\right)^\dagger|\mathrm{g},0\rangle\equiv\hbar\omega_\lambda$ for the diagonal field energy matrix elements.

\subsubsection{Results for the dipolar $\hat{\mathbf{A}}\cdot\hat{\mathbf{p}}$ coupling \label{subsubsec:AdotPDipoleAppdx}}  

For the $\hat{\mathbf{A}}\cdot\hat{\mathbf{p}}$ coupling in the dipole approximation (\ref{eq:DipoleFacchiCoupling}) we obtain
    \begin{multline} \label{eq:IntMatrixElementDF}
\langle\mathrm{e},0|\hat{H}_I^R \ \left(\hat{a}_{\mathrm{avg}}^\lambda\right)^\dagger|\mathrm{g},0\rangle=\frac{\mathrm{1}}{2\pi}\sqrt{\frac{\hbar c}{3\epsilon_0}}\left|\mathbf{d}_{\mathrm{ge}}\right| \ \omega_{\mathrm{eg}} \\
\left(\frac{k_{\lambda+1}^2-k_\lambda^2}{2}\right)^{-\frac{1}{2}}
    \end{multline}
with $\hat{a}_{\mathrm{avg}}^\lambda\equiv\hat{a}_{\mathrm{avg}}^{k_\lambda k_{\lambda+1}}$ defined through
\begin{equation} \label{eq:AverageAAPD}
    \mathrm{i}\left(2\pi\right)^{-\frac{3}{2}} \  \ \int_{k_\lambda}^{k_{\lambda+1}}\mathrm{d}k \ k^{\frac{3}{2}} \ \hat{a}_{\mathrm{sph}}\left(k\right)\equiv\hat{a}_{\mathrm{avg}}^\lambda
    \left(\int_{k_\lambda}^{k_{\lambda+1}}\mathrm{d}k \ k\right)^{\frac{1}{2}}
\end{equation}
 and
        \begin{equation} \label{eq:FieldMatrixElementDF}
          \langle\mathrm{g},0|\hat{a}_{\mathrm{avg}}^\lambda\ \hat{H}_R\ \hat{a}_{\mathrm{avg}}^{\zeta\dagger}|\mathrm{g},0\rangle=\frac{2}{3}\ \hbar c \ \delta_{\lambda\zeta} \ \frac{k_{\lambda+1}^3-k_\lambda^3}{k_{\lambda+1}^2-k_\lambda^2}.
        \end{equation}
        The matrix elements (\ref{eq:IntMatrixElementDF}) and (\ref{eq:FieldMatrixElementDF}) are the ones we use for our numerical calculations. In the body of the article we denote them as $\langle\mathrm{e},0|\hat{H}_I^R\,\left(\hat{a}_{\mathrm{avg}}^\lambda\right)^\dagger|\mathrm{g},0\rangle\equiv-\mathrm{i}\hbar G_\lambda^*\left(\mathbf{r}_0\right)$ for the coupling matrix element and $\langle\mathrm{g},0|\hat{a}_{\mathrm{avg}}^\lambda\ \hat{H}_R\ \left(\hat{a}_{\mathrm{avg}}^\lambda\right)^\dagger|\mathrm{g},0\rangle\equiv\hbar\omega_\lambda$ for the diagonal field energy matrix elements.
        
\subsubsection{Results for the exact $\hat{\mathbf{A}}\cdot\hat{\mathbf{p}}$ coupling \label{subsubsec:AdotPExactAppdx}} 
Finally for the exact $\hat{\mathbf{A}}\cdot\hat{\mathbf{p}}$ coupling (\ref{eq:FacchiCoupling}) we obtain
    \begin{multline} \label{eq:IntMatrixElementF}
\langle\mathrm{e},0|\hat{H}_I^R\ \left(\hat{a}_{\mathrm{avg}}^\lambda\right)^\dagger|\mathrm{g},0\rangle=\frac{\mathrm{1}}{2\pi} \ \sqrt{\frac{\hbar c}{3\epsilon_0}} \ \left|\mathbf{d}_{\mathrm{ge}}\right| \ \omega_{\mathrm{eg}}\\
\left(I^{\left(2\right)}\left(k_{\lambda+1}\right)-I^{\left(2\right)}\left(k_\lambda\right)\right)^{-\frac{1}{2}}
    \end{multline}
where
    \begin{equation} \label{eq:I2}
      I^{\left(2\right)}\left(k\right)\equiv-\frac{k_{\mathrm{X}}^2}{6} \ \frac{1}{\left[1+\left(\frac{k}{k_{\mathrm{X}}}\right)^2\right]^3}
    \end{equation}
with $\hat{a}_{\mathrm{avg}}^\lambda\equiv\hat{a}_{\mathrm{avg}}^{k_\lambda k_{\lambda+1}}$ defined through 
\begin{multline} \label{eq:AverageAAP}
    \mathrm{i}\left(2\pi\right)^{-\frac{3}{2}} \  \ \int_{k_\lambda}^{k_{\lambda+1}}\mathrm{d}k\ \frac{k^{\frac{3}{2}}}{\left[1+\left(\frac{k}{k_{\mathrm{X}}}\right)^2\right]^2} \ \hat{a}_{\mathrm{sph}}\left(k\right)\\
    \equiv\hat{a}_{\mathrm{avg}}^\lambda\left(\int_{k_\lambda}^{k_{\lambda+1}}\mathrm{d}k\ \frac{k}{\left[1+\left(\frac{k}{k_{\mathrm{X}}}\right)^2\right]^4}\right)^{\frac{1}{2}}.
\end{multline}
We also find
        \begin{equation} \label{eq:FieldMatrixElementF}
          \langle\mathrm{g},0|\hat{a}_{\mathrm{avg}}^\lambda\,\hat{H}_R\,\hat{a}_{\mathrm{avg}}^{\zeta\dagger}|\mathrm{g},0\rangle=\hbar c\ \delta_{\lambda\zeta} \ \frac{I^{\left(3\right)}\left(k_{\lambda+1}\right)-I^{\left(3\right)}\left(k_\lambda\right)}{I^{\left(2\right)}\left(k_{\lambda+1}\right)-I^{\left(2\right)}\left(k_\lambda\right)}
        \end{equation}
        where
\begin{widetext}
        \begin{equation} \label{eq:I3}
          \begin{split}
            I^{\left(3\right)}\left(k_{\lambda+1}\right)-I^{\left(3\right)}\left(k_\lambda\right)&=\int_{k_\lambda}^{k_{\lambda+1}}\mathrm{d}k\frac{k^2}{\left[1+\left(\frac{k}{k_{\mathrm{X}}}\right)^2\right]^4}\\
            &=\frac{k_{\mathrm{X}}^3}{48}\left[\frac{k_\lambda k_{\mathrm{X}} \left(k_{\mathrm{X}}^2-3 k_\lambda^2\right) \left(k_\lambda^2+3 k_{\mathrm{X}}^2\right)}{\left(k_\lambda^2+k_{\mathrm{X}}^2\right)^3}\frac{k_{\lambda+1} k_{\mathrm{X}} \left(k_{\mathrm{X}}^2-3 k_{\lambda+1}^2\right) \left(k_{\lambda+1}^2+3 k_{\mathrm{X}}^2\right)}{\left(k_{\lambda+1}^2+k_{\mathrm{X}}^2\right)^3}\right.\\
            &\left.-3\arctan\left(\frac{k_\lambda}{k_{\mathrm{X}}}\right)+3\arctan\left(\frac{k_{\lambda+1}}{k_{\mathrm{X}}}\right)\right].
          \end{split}
        \end{equation}
\end{widetext}
        The matrix elements (\ref{eq:IntMatrixElementF}) and (\ref{eq:FieldMatrixElementF}) are the ones we use for our numerical calculations. In the body of the article we denote them as $\langle\mathrm{e},0|\hat{H}_I^R\,\left(\hat{a}_{\mathrm{avg}}^\lambda\right)^\dagger|\mathrm{g},0\rangle\equiv-\mathrm{i}\hbar G_\lambda^*\left(\mathbf{r}_0\right)$ for the coupling martix element and $\langle\mathrm{g},0|\hat{a}_{\mathrm{avg}}^\lambda\,\hat{H}_R\,\left(\hat{a}_{\mathrm{avg}}^\lambda\right)^\dagger|\mathrm{g},0\rangle\equiv\hbar\omega_\lambda$ for the diagonal field energy matrix elements.

\begin{acknowledgments}
Vincent Debierre acknowledges support from CNRS (INSIS doctoral grant). Thomas Durt acknowledges support from the COST 1006 and COST 1043 actions. Edouard Brainis acknowledges the Belgian Science Policy Office (IAP 7.35 photonics@be) and the Horizon2020 program (MSCA-ETN phonsi) for funding this research. 
\end{acknowledgments}

\bibliography{Biblio}
\end{document}